\documentclass[11pt]{article}
\usepackage[a4paper,margin=1in]{geometry}
\usepackage{iftex}

\ifPDFTeX
  \usepackage[T1]{fontenc}
  \usepackage[utf8]{inputenc}
  \usepackage{lmodern}
\else
  \usepackage{fontspec}
  \usepackage{xeCJK}

\fi

\usepackage[numbers,sort&compress]{natbib}
\usepackage{amsmath,amssymb,amsfonts,mathtools}
\usepackage{bm}
\usepackage{graphicx}
\usepackage{caption}
\usepackage{booktabs}
\usepackage{enumitem}
\usepackage{microtype}
\usepackage{url}
\usepackage[hidelinks]{hyperref}
\usepackage{xspace}

\setlist[itemize]{leftmargin=1.4em}
\newcommand{\system}{Vi-Liquid\xspace}

\title{Smartphone-Based Identification of Unknown Liquids via Active Vibration Sensing}
\author{
Yongzhi Huang\\
Information Hub, Data Science and Analytics (DSA) Thrust\\
The Hong Kong University of Science and Technology (Guangzhou)\\
Guangzhou, China\\
\texttt{huangyongzhi@email.szu.edu.cn}
}
\date{}

\begin{document}

\maketitle

\begin{abstract}
Traditional liquid-analysis instruments are expensive, invasive, and rarely available to end users. This paper studies whether a commodity smartphone can identify an unknown liquid without external sensing hardware. Our key observation is that, under fixed mechanical excitation, liquid viscosity induces both a measurable boundary shear load and a dissipation pattern on the container wall. Based on this observation, we build a physics-grounded vibration--viscosity model that links steady-state amplitude and free-decay attenuation to viscosity, and we realize the model on a smartphone using only the built-in vibro-motor and accelerometer. A practical deployment must overcome three system constraints: severe under-sampling at the mobile operating-system API, strong straight-path self-interference from the motor to the inertial sensor, and volume-dependent changes in the coupled wall--liquid resonance. \system addresses them with phase-shift-based supersampling rate reconstruction, OMP-based sparse recovery, spectral subtraction of a calibrated straight-path template, and volume compensation in the frequency domain. Across 30 liquids, \system achieves a mean relative viscosity error of 2.9\% and 95.47\% identification accuracy, and it also supports proof-of-concept screening for water contamination, changes in urine composition, and alcohol concentration. These results show that liquid identification by physics-based active vibration sensing is feasible on commodity smartphones without specialized sensor add-ons or liquid-specific training.
\end{abstract}


\begin{figure}[tbp]
\centering
\includegraphics[width=\linewidth]{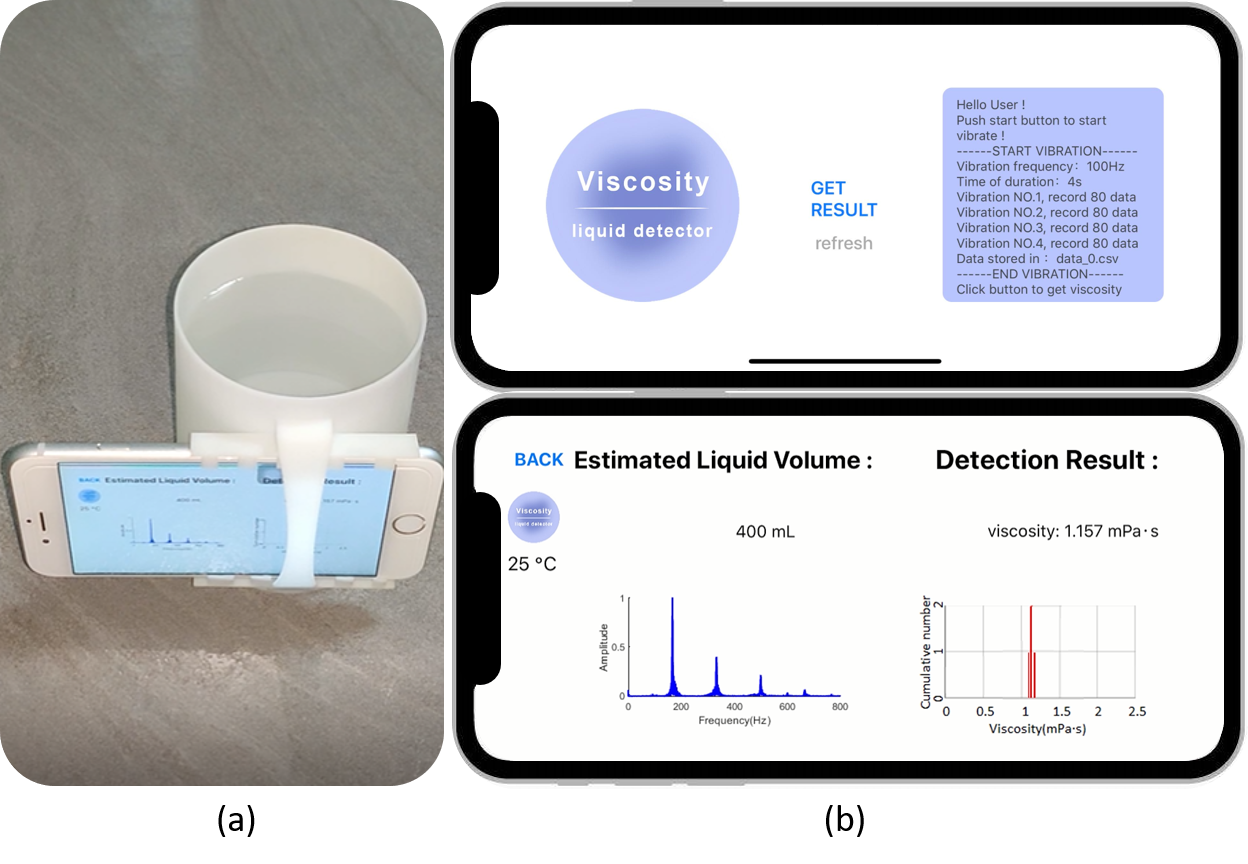}
\caption{(a) Using an iPhone 7 to detect a liquid. (b) The user interface of \system.}
\label{fig:phone-detection}
\end{figure}

\section{Introduction}

Liquid testing is useful in safety inspection, beverage authentication, nutritional tracking, household water screening, and home health monitoring. Most ubiquitous liquid-testing systems, however, still depend on external hardware such as RFID readers, dedicated photodiodes, UWB radios, or photoacoustic attachments \cite{dhekne2018liquid,rahman2016nutrilyzer}. More broadly, recent mobile and ubiquitous sensing systems have shown that commodity devices can recover subtle physical variables from weak RF, vibration, optical, and magnetic channels \cite{shuxin2017wi,yongzhi2018mm,yongzhi2018oinput,qianru2022magear,chen2023lit}. This trend motivates a natural question: can a commodity smartphone identify an unknown liquid without external sensors?

The answer is not obvious. Existing smartphone-oriented liquid systems are either modality-specific or require prior knowledge about the liquid. CapCam \cite{yue2019liquid} uses the phone's vibro-motor, flashlight, and camera to measure surface tension, but it needs prior information about transparent liquids and does not work well for opaque liquids. Subsequent optical systems target liquor quality monitoring or beverage deterioration \cite{huang2021lili,yongzhi2023beverage}, and a patent on structured-light capillary-wave measurement further shows the promise of optical smartphone liquid sensing \cite{wu2020structlight}. Yet these approaches still rely on optical access and are not designed to infer viscosity solely from inertial sensing.

In this paper, we focus on \emph{viscosity}, an intrinsic property of fluids that directly governs resistance to relative molecular motion. Viscosity is not a universal one-to-one identifier for every conceivable liquid. Still, within a bounded candidate set and under controlled temperature and container conditions, it is a powerful discriminative feature and directly useful for concentration estimation and anomaly screening. Conventional viscometers \cite{tripolino2017body,de2004shear,blom1984torsion,shih2001simultaneous,riesch2008characterizing,gottlieb1979zero} are expensive, invasive, and impractical for everyday use. Our goal is therefore narrower and more realistic than chemistry-grade liquid analysis: we seek smartphone-based viscosity estimation and candidate-set liquid identification using only the phone's built-in actuation and inertial sensing.

\paragraph{Approach.}
We present \system, a lightweight smartphone system that estimates liquid viscosity using only the built-in linear resonant actuator and accelerometer. As shown in Figure \ref{fig:phone-detection}, the user attaches the smartphone to the container and starts the app. The phone generates active vibration, collects the reflected vibration through the accelerometer, and then feeds the signal into a vibration--viscosity estimator. The central insight is that the liquid produces a viscosity-dependent resistance when the container wall oscillates. That resistance changes both the amplitude of the steady-state response and the decay rate after actuation stops. By measuring both effects, the system can infer viscosity from commodity motion sensors.

\paragraph{Challenges.}
Turning this idea into a usable smartphone system is non-trivial. First, we need a physical model that relates liquid viscosity to the vibration measured on the phone. Second, the sensing stack is heavily constrained by the mobile OS: the built-in accelerometer can sample only up to 100~Hz via the public API, whereas the vibromotor resonates near 167~Hz. Using the raw signal directly, therefore, causes severe aliasing and peak distortion. Third, practical use introduces strong confounders. The dominant one is straight-path self-interference, where the motor couples directly to the accelerometer and overwhelms the much weaker liquid-reflected signal. Another is volume change, which alters the coupled wall--liquid resonance and causes the same liquid to produce different spectral envelopes at different fill levels.

\paragraph{Our response.}
To address the first challenge, we derive a reduced-order single-degree-of-freedom model for the coupled wall--liquid system. The model shows that the steady-state amplitude depends on both the viscosity-induced shear load and the effective damping, while the decaying stage isolates the damping term. This leads to a closed-form estimator that uses both stages, which is why \system deliberately excites the phone in bursts rather than with one uninterrupted tone. To address the second challenge, we propose \emph{Supersampling Rate Reconstruction} (SRR), which leverages phase diversity across repeated bursts to fold multiple low-rate measurements into a higher-resolution waveform, and then refines it using OMP-based sparse recovery. To address the third challenge, we explicitly model and subtract straight-path interference in the frequency domain and compensate volume-dependent spectral changes with a calibrated weighting scheme.

\paragraph{Results.}
Across 30 liquids, \system achieves a mean relative viscosity error of 2.9\% and identifies liquids with an average accuracy of 95.47\%. It also distinguishes highly similar beverages such as Coca-Cola and Pepsi, differentiates contamination-relevant water samples, estimates sodium urate and protein concentration in synthetic urine, and measures alcohol concentration in the tested range. A preliminary conference version and a later journal extension of this research line appear in \cite{huang2021vi,huang2021portable}, and this arXiv-oriented manuscript refines the technical presentation, corrects several modeling ambiguities, and more explicitly connects the experiments to the underlying theory.

\paragraph{Contributions.}
The main contributions of this work are:
\begin{itemize}
  \item We introduce a smartphone-based liquid sensing system that uses only built-in actuation and inertial sensing to estimate viscosity, enabling candidate-set liquid identification and concentration screening without external sensors or large liquid-specific training sets.
  \item We derive a physics-grounded vibration--viscosity model and show that viscosity becomes identifiable only when steady-state amplitude and decay-stage attenuation are jointly measured. This analysis explains the sensing protocol rather than merely post hoc data fitting.
  \item We design a practical signal-processing pipeline for smartphones, including SRR, OMP-based waveform refinement, straight-path interference subtraction, and volume compensation, and validate the system through comprehensive experiments.
\end{itemize}

The rest of the paper is organized as follows. Section~2 presents the vibration model and the closed-form estimator. Section~3 uses standalone hardware to validate the modeling assumptions. Section~4 gives the system overview. Section~5 explains the smartphone realization. Section~6 evaluates the end-to-end system. Section~7 reviews related work and the broader sensing context. Section~8 concludes and discusses limitations and future directions.

\begin{figure}[tbp]
\centering
\includegraphics[width=0.8\linewidth]{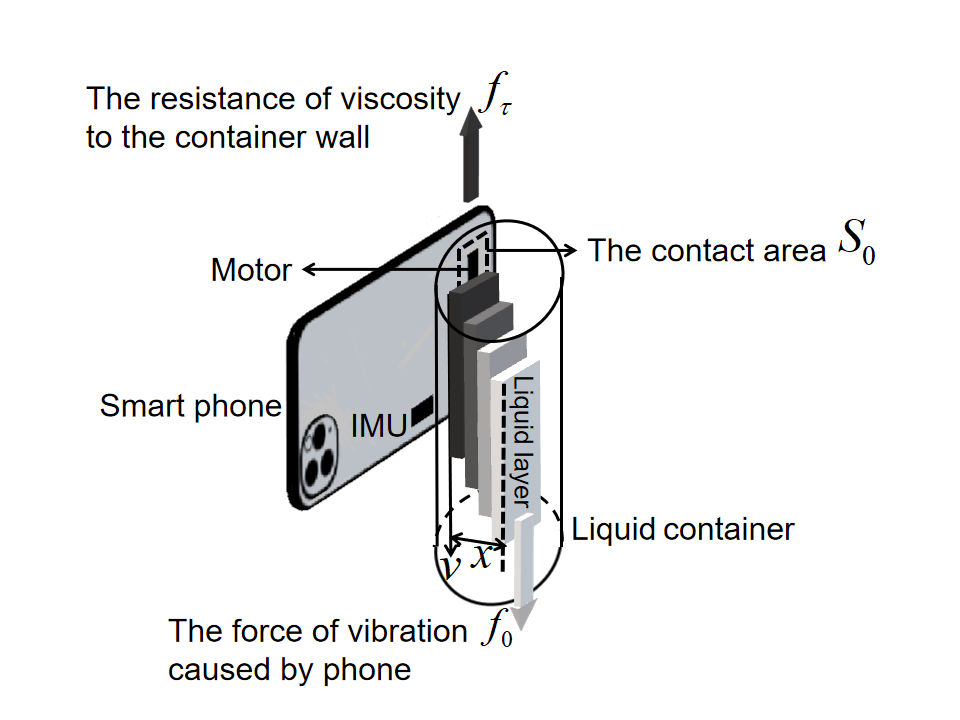}
\caption{Physical measurement model of \system.}
\label{fig:vibration-system}
\end{figure}

\section{Vibration Model for Viscosity Measurement}

This section develops a reduced-order measurement model. The goal is not to reproduce the full fluid dynamics of the liquid but to preserve the dominant terms that determine the vibration observed by the smartphone.

\subsection{Liquid Viscosity and the Shearing Force}

Viscosity is an essential physical property of liquids. At the microscopic scale, molecules must overcome an activation barrier to rearrange under shear. In Eyring's activated-process view, dynamic viscosity can be expressed as \cite{eyring1935activated,chapman1990mathematical}
\begin{equation}
\eta = \frac{hN_A}{V_m}\exp\left(\frac{\Delta G^\ddagger}{RT}\right),
\end{equation}
where $V_m$ is the molar volume, $\Delta G^\ddagger$ is the activation free energy, $T$ is the absolute temperature, $h$ is Planck's constant, $N_A$ is Avogadro's constant, and $R$ is the gas constant. This equation is important mainly for intuition: liquids with different molecular structures or concentrations generally have different activation barriers and hence different viscosities. Throughout the experiments, we maintain a stable ambient temperature, so $T$ can be treated as constant.

At the macroscopic scale, a Newtonian liquid obeys the constitutive relation
\begin{equation}
\tau = \eta \frac{\partial u}{\partial y},
\end{equation}
where $\tau$ is shear stress and $\partial u/\partial y$ is the velocity gradient normal to the wall. For the near-wall oscillation induced by the smartphone, we approximate the local velocity profile as linear over an effective shear depth $d$. The resulting shear force on the container wall is
\begin{equation}
f_\tau = \eta S_0 \frac{v}{d},
\label{fandviscosity}
\end{equation}
where $S_0$ is the effective wall--liquid contact area and $v$ is the characteristic velocity of the oscillating wall. For a fixed container, phone placement, volume, and actuation frequency, $S_0$, $v$, and $d$ vary little within one measurement. We therefore absorb them into a constant $\gamma = S_0 v/d$ and obtain the linear relation $f_\tau = \gamma \eta$.

Equation~\ref{fandviscosity} is the bridge between fluid physics and sensing: it says that, under controlled geometry and actuation, viscosity manifests as a measurable boundary shear load.

\subsection{Viscosity Calculation Leveraging Vibration}

During a measurement burst, the vibration can be separated into three stages: a startup transient, a steady forced stage, and a free-decay stage after the motor pauses. We discard the startup transient because motor spin-up is highly dynamic and weakly repeatable, and the latter two stages are much more stable and jointly identify viscosity.

We model the wall-coupled near-wall liquid layer as an effective single-degree-of-freedom oscillator. Let $x(t)$ denote the wall displacement measured along the actuation axis. During steady forcing,
\begin{equation}
m\frac{\mathrm{d}^2x}{\mathrm{d}t^2} + \beta \frac{\mathrm{d}x}{\mathrm{d}t} + kx = (f_0 - f_\tau)\sin(\omega t),
\label{differential}
\end{equation}
where $m$ is the effective hydrodynamic mass of the oscillating near-wall liquid, $\beta$ is the effective damping coefficient, $k$ is the effective stiffness of the phone--container structure, $f_0$ is the motor-force amplitude, and $\omega$ is the actuation angular frequency. Note that $m$ is \emph{not} the total liquid mass, and it represents only the portion of liquid that is dynamically coupled to the wall.

The steady-state particular solution of Equation~\ref{differential} is
\begin{equation}
x_{vib}(t) = \frac{f_0-f_\tau}{\sqrt{(k-\omega^2 m)^2+(\beta\omega)^2}}\sin(\omega t-\phi),
\label{xs}
\end{equation}
where
\[
\phi = \arctan\left(\frac{\beta \omega}{k-m\omega^2}\right).
\]
Therefore, the steady-state amplitude is
\begin{equation}
A = \frac{f_0-f_\tau}{\sqrt{(k-\omega^2 m)^2+(\beta\omega)^2}}.
\label{A}
\end{equation}

Equation~\ref{A} reveals an identifiability problem: the amplitude is affected by both the shear load $f_\tau$ and the damping $\beta$. Measuring $A$ alone is therefore insufficient.

When actuation stops, the system enters free decay:
\begin{equation}
m\frac{\mathrm{d}^2x}{\mathrm{d}t^2} + \beta \frac{\mathrm{d}x}{\mathrm{d}t} + kx = 0.
\end{equation}
For the viscosity range in which the response remains oscillatory, the solution is
\begin{equation}
x_{decay}(t) = A_0 e^{-\frac{\beta}{2m}t}\sin(\omega_d t + \theta),
\end{equation}
where $\omega_d = \sqrt{k/m-(\beta/2m)^2}$ is the damped natural frequency. If $T_d = 2\pi/\omega_d$ denotes the period of adjacent peaks in the decaying stage, then the logarithmic decrement between adjacent peaks is
\begin{equation}
\Lambda = \frac{x_{decay}(t)}{x_{decay}(t+T_d)} = e^{T_d \cdot \frac{\beta}{2m}}.
\label{reduction_factor}
\end{equation}
Hence
\[
\beta = \frac{2m}{T_d}\ln \Lambda.
\]

Substituting $\beta$ into Equation~\ref{A} yields
\[
f_\tau = f_0 - A\sqrt{(k-\omega^2 m)^2+(\beta\omega)^2}.
\]
Finally, from Equation~\ref{fandviscosity},
\[
\eta = \frac{d}{S_0 v}\left[f_0 - A\sqrt{(k-\omega^2 m)^2+(\beta\omega)^2}\right].
\]
This closed-form estimator explains the whole sensing protocol: the decaying stage identifies $\beta$, and the steady stage then identifies $f_\tau$ and therefore $\eta$.

The model relies on several assumptions consistent with the target operating regime: small-amplitude wall motion, fixed container geometry during a single trial, negligible bulk sloshing, stable temperature, and no gross phone motion. Section~3 validates these assumptions experimentally before we move to the smartphone implementation.

\begin{figure}[tbp]
\centering
\includegraphics[width=\linewidth]{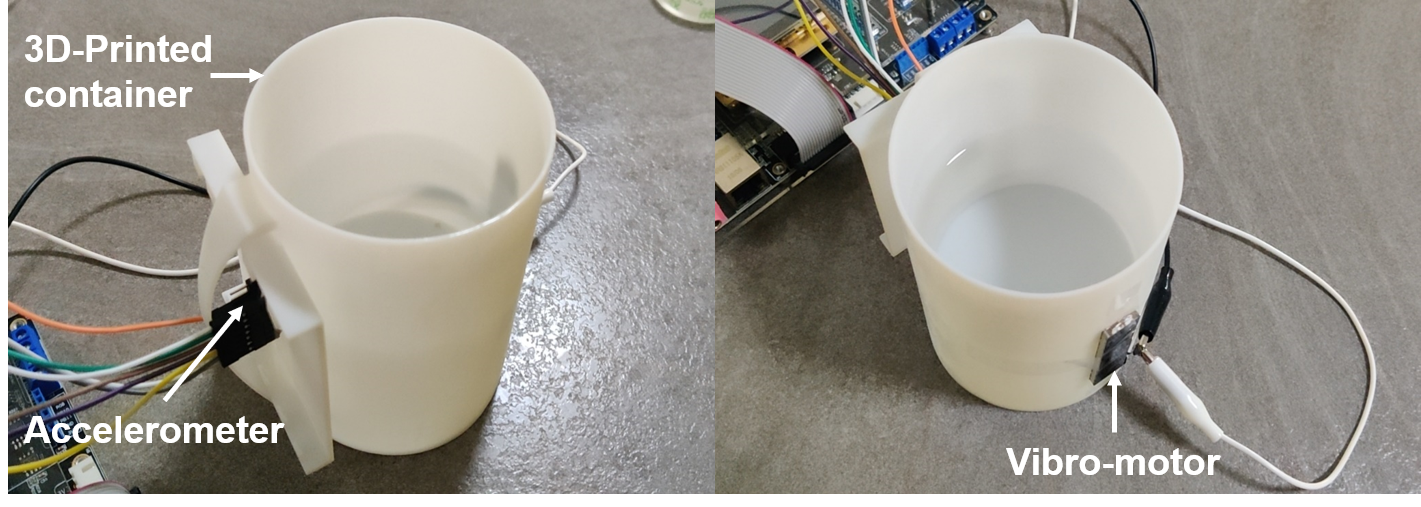}
\caption{Feasibility-study setup with a separate motor, accelerometer, container, and alternating-current power supply.}
\label{fig:single-motor}
\end{figure}

\begin{figure}[tbp]
\centering
\includegraphics[width=0.85\linewidth]{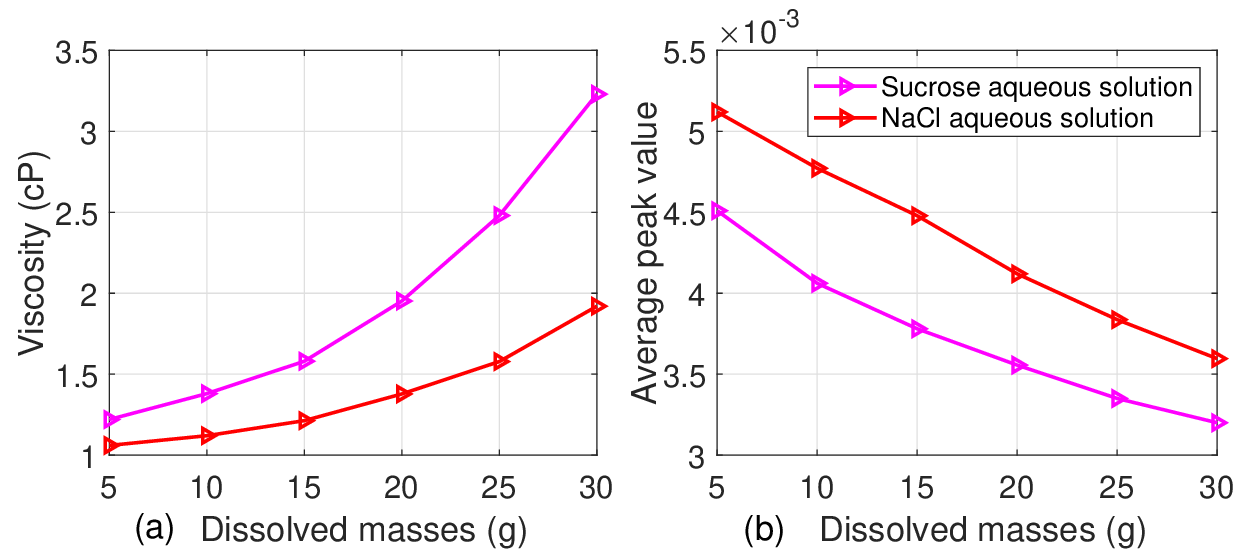}
\caption{Dissolving different masses of sucrose and NaCl in solution. (a) Change of viscosity. (b) Change of average peak value.}
\label{fig:Mass-viscosity}
\end{figure}

\begin{figure}[tbp]
\centering
\includegraphics[width=0.75\linewidth]{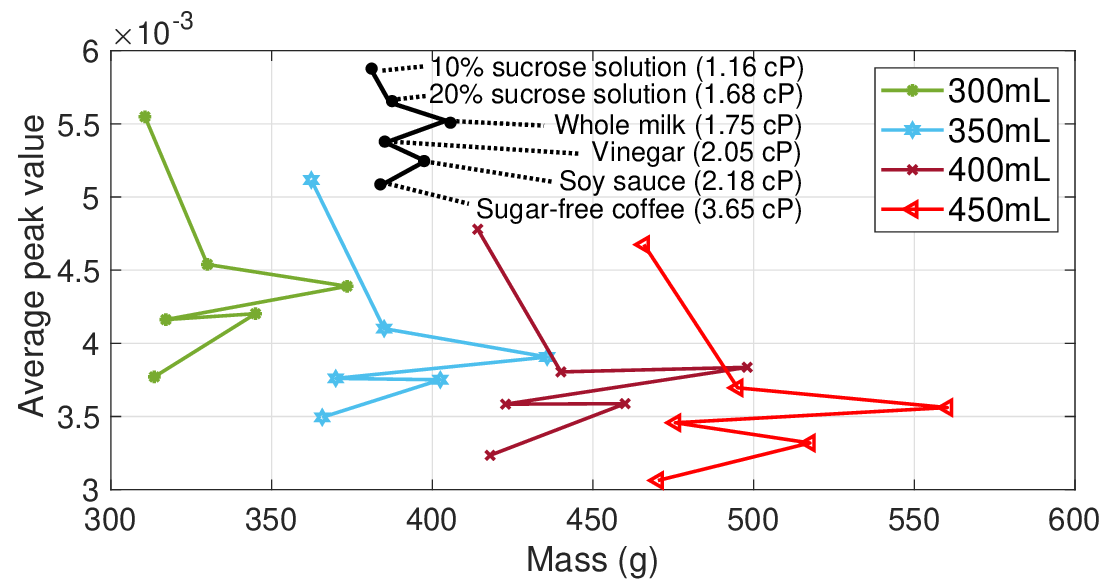}
\caption{Distribution of different volumes of liquid in the mass--amplitude plane.}
\label{fig:distribution-different-volumes-liquid-mass-amplitude}
\end{figure}

\begin{figure}[tbp]
\centering
\includegraphics[width=0.55\linewidth]{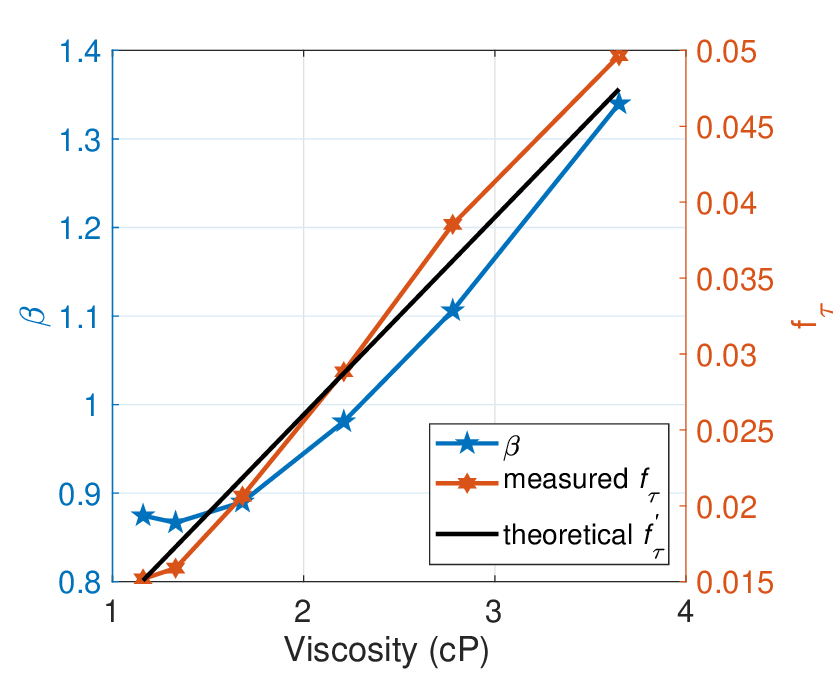}
\caption{Measured $\beta$ and $f_\tau$, together with theoretical $f'_\tau$.}
\label{fig:beta_f_tau}
\end{figure}

\section{Feasibility Study}

This section validates the reduced-order model using standalone hardware. The purpose is to isolate the physics before confronting the sensing limitations of a smartphone.

\subsection{Experimental Setup}

We used a Form~3 3D printer to fabricate a 500~mL tableware-resin cup with two side slots aligned at the 250~mL height mark. An iPhone~7 Taptic Engine and a BMI160 three-axis accelerometer were mounted in the slots, as shown in Figure~\ref{fig:single-motor}. The motor vibrated at 167~Hz, matching the phone used later in the end-to-end system. In the feasibility study, the standalone accelerometer was sampled at 1600~Hz to produce high-resolution waveforms. Because we estimate viscosity from shear load in the actuation direction, we analyze only the $x$-axis readings.

Ground truth was obtained with an $ATAGO\text{-}VISCO^{TM}\ 895$ rotary viscometer, whose measurement range is 1 to $3.5\times 10^{8}$~cP with 0.01~cP resolution and approximately 1\% relative error. Unless otherwise stated, all experiments were run in a laboratory at 25$^\circ$C. This high-rate standalone setup is used only to validate the model, and the smartphone system in Section~6 uses the built-in sensor at the API-limited rate.

\subsection{The Viscosity Uniqueness of Liquids}

We first study whether viscosity changes are visible in the received vibration. To isolate viscosity from geometric effects, we need two solutions whose volumes increase at similar rates under equal added mass but whose viscosities increase differently. In pilot trials, sucrose solution and NaCl solution satisfied this requirement at 18$^\circ$C.

We dissolved increasing masses of sucrose and NaCl into 450~mL of water and recorded both viscometer readings and vibration signals. Figure~\ref{fig:Mass-viscosity}(a) shows that equal added mass does \emph{not} imply equal viscosity change: sucrose thickens the solution much more strongly than NaCl. Figure~\ref{fig:Mass-viscosity}(b) shows a corresponding drop in average peak amplitude. This is exactly what Equation~\ref{A} predicts: larger viscosity increases both the shear load $f_\tau$ and the effective damping $\beta$, both of which reduce the steady-state amplitude. Importantly, even the smallest viscosity gap in the figure, about 0.16~cP, still causes a visible amplitude difference.

At the same time, Figure~\ref{fig:Mass-viscosity}(b) also makes clear why amplitude alone is insufficient. The average peak value depends on more than viscosity, which motivates the next experiments on mass, volume, and decay-stage attenuation.

\subsection{The Impact of Mass}

Does the \emph{total} mass of liquid affect the vibration response? The model says that it should not be the primary variable because the effective mass $m$ in Equation~\ref{differential} is the near-wall oscillating layer rather than the full bulk volume. For oscillatory motion, the velocity penetrates only a Stokes boundary layer of thickness
\[
\delta = \sqrt{\frac{2\eta}{\rho\omega}},
\]
so the coupled liquid mass is approximately $m \approx \rho S_0 \delta$. At 167~Hz, this boundary layer is only on the order of tens to hundreds of micrometers for the low- and medium-viscosity liquids in our study, which is far smaller than the full liquid depth.

We therefore prepared six liquids with different viscosities (10\% and 20\% sucrose solutions, whole milk, vinegar, soy sauce, and sugar-free coffee) at four volumes, for a total of 24 samples, and plotted the average peak amplitude versus total mass in Figure~\ref{fig:distribution-different-volumes-liquid-mass-amplitude}. The samples do not align by total mass. For example, at the same volume, whole milk can have a higher total mass but a lower amplitude than a lighter sucrose solution because its viscosity is higher. This directly supports the boundary-layer interpretation: the response is governed by the near-wall coupled layer rather than by the total liquid mass in the container.

\subsection{The Volume Interference}

Volume is different. Increasing the fill level changes both the effective wall--liquid contact area and the dynamic loading of the container, so the response amplitude and spectral envelope both change. Through Equation~\ref{fandviscosity}, a larger effective contact area increases the shear load, which tends to reduce amplitude. At the same time, the extra liquid alters the resonance of the coupled system, so the effect is not a simple linear scaling.

\begin{figure}[tbp]
\centering
\includegraphics[width=0.62\linewidth]{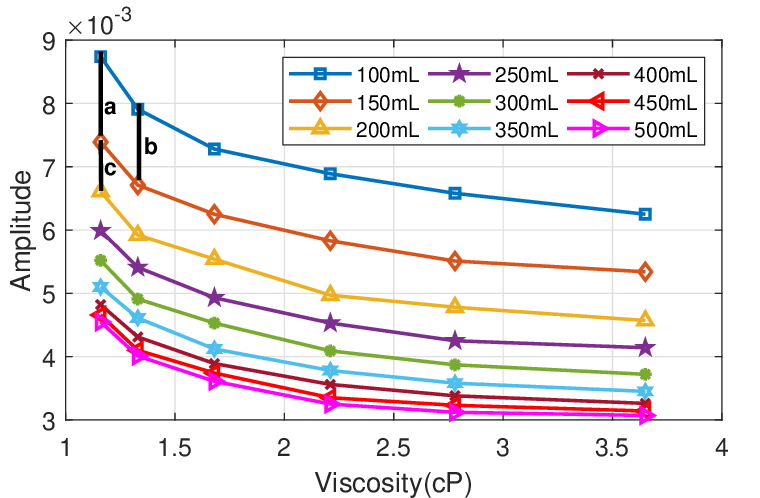}
\caption{The attenuation of volume to viscosity is not proportional.}
\label{fig:volume_attenuation_not_proportional}
\end{figure}

To quantify the effect, we prepared six sucrose solutions with mass concentrations of 10\%, 15\%, 20\%, 25\%, 30\%, and 35\%, corresponding to viscosities of 1.16, 1.33, 1.68, 2.21, 2.78, and 3.65~cP. We then filled the cup from 100~mL to 500~mL, collected 50 measurements at 50~mL increments, and computed the average peak amplitude. Figure~\ref{fig:volume_attenuation_not_proportional} shows two important properties. First, amplitude decreases as volume increases, as expected. Second, the decay with volume is clearly \emph{non-linear}: the spacing between curves shrinks at larger volumes, and different viscosities can exhibit similar attenuation slopes. This is why a single scalar normalization is not enough, and the compensation in Section~5.6 must operate in the frequency domain.

\subsection{Attenuation in Decaying-State}

\begin{figure}[tbp]
\centering
\includegraphics[width=0.66\linewidth]{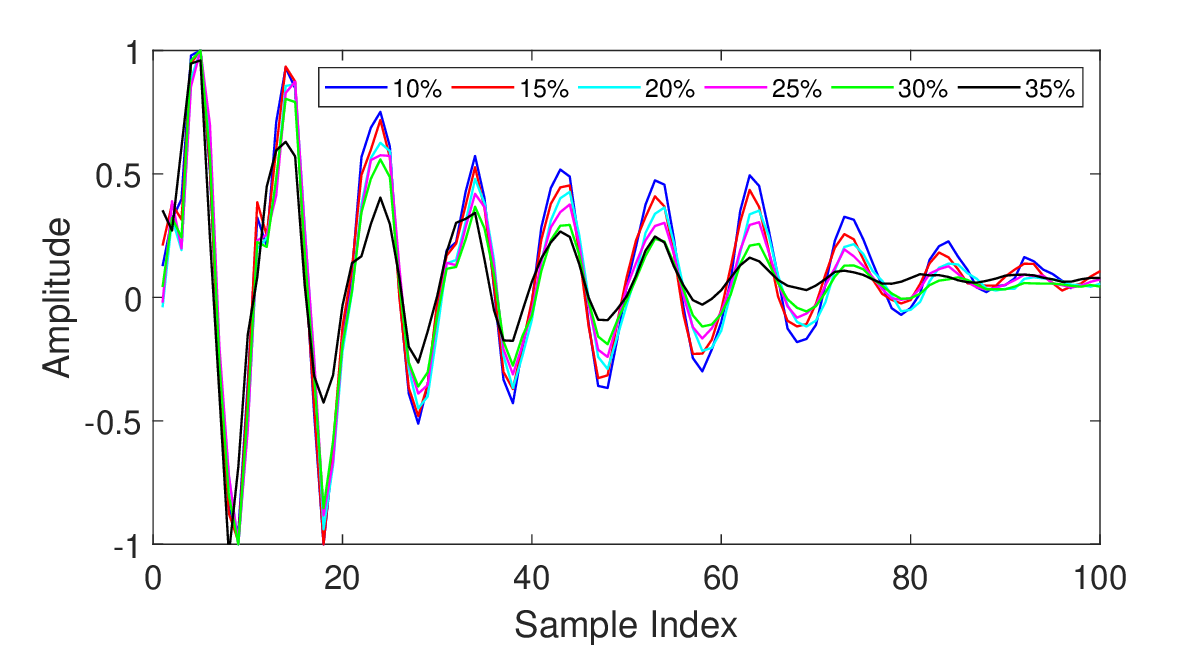}
\caption{Decaying-state signals of solutions with different concentrations after the motor stops.}
\label{fig:motor_stop}
\end{figure}

Next, we verify the second half of the model: the decay stage should reveal damping. We filled the container with 400~mL sucrose solution at six concentrations (10\%, 15\%, 20\%, 25\%, 30\%, and 35\%). As shown in Figure~\ref{fig:motor_stop}, once the motor stops, the vibration decays more rapidly for more viscous liquids. This directly matches Equation~\ref{reduction_factor}: larger $\beta$ yields faster exponential attenuation.

We estimate the mean logarithmic decrement $\bar{\Lambda}$ from adjacent peaks, solve for $\beta$, and then recover $f_\tau$ and $\eta$. In Figure~\ref{fig:beta_f_tau}, the theoretical $f'_\tau$ is computed from the viscometer ground truth through Equation~\ref{fandviscosity}, and the measured $f_\tau$ is computed from the vibration data. The two agree closely, with a mean relative error of 4.57\%. This experiment closes the loop between the theory in Section~2 and the sensing signal: the decaying stage is not just qualitatively different across liquids, it quantitatively recovers the damping needed by the estimator.

\begin{figure}[tbp]
\centering
\includegraphics[width=0.62\linewidth]{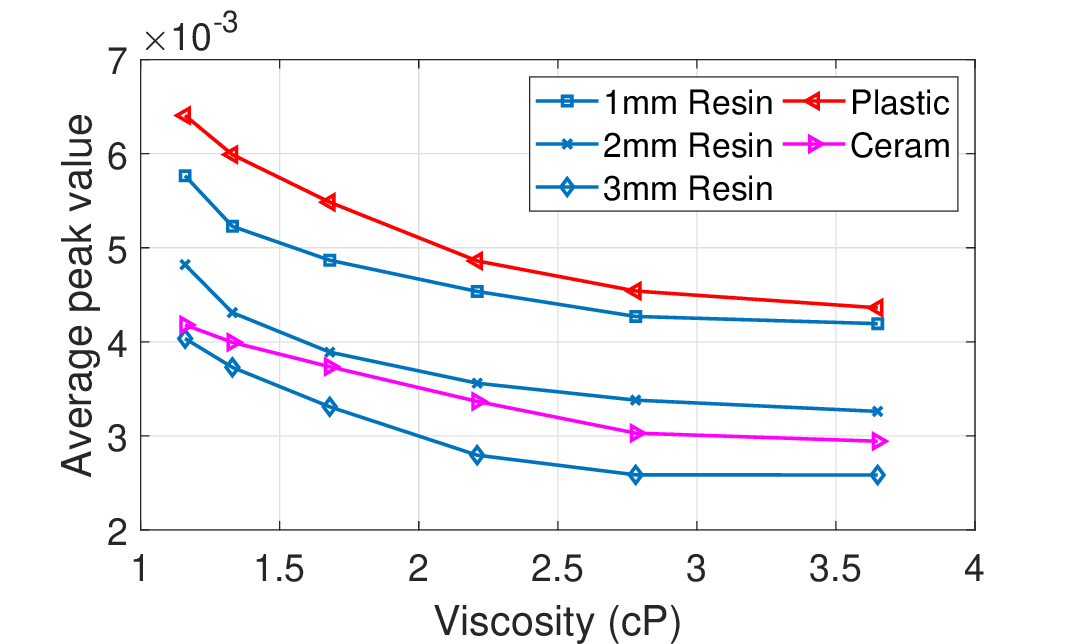}
\caption{Average peak value under different container materials and thicknesses.}
\label{fig:material-thickness}
\end{figure}

\begin{figure}[tbp]
\centering
\includegraphics[width=\linewidth]{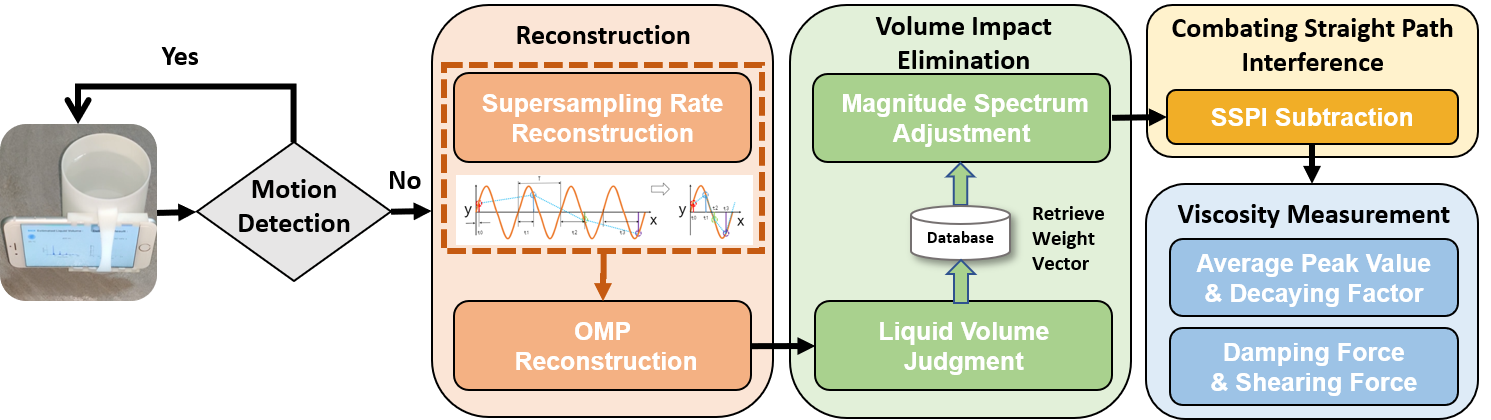}
\caption{The workflow of \system.}
\label{fig:system-overview}
\end{figure}

\begin{figure}[tbp]
\centering
\includegraphics[width=0.85\linewidth]{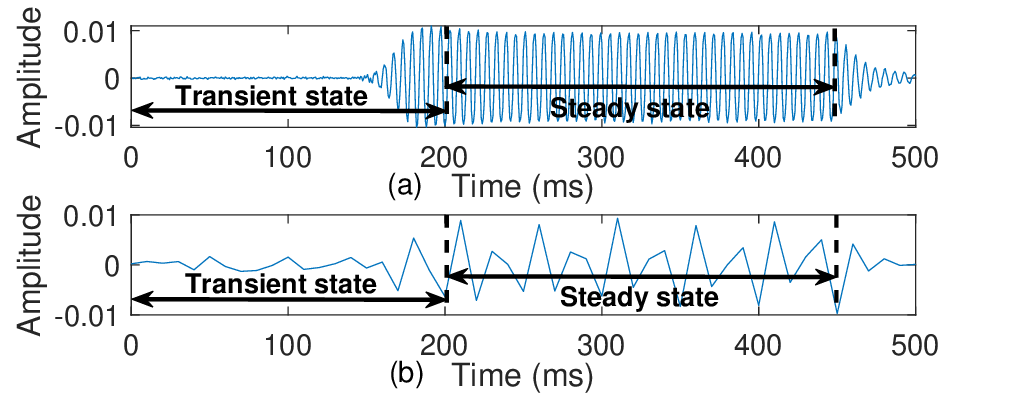}
\caption{Transient signals at different sampling rates. (a) 1600~Hz. (b) 100~Hz.}
\label{fig:transient-signal}
\end{figure}

\subsection{The Impact of Container}

The container is part of the mechanical system, so its material and thickness affect the effective stiffness $k$ and, consequently, the transfer function. We evaluated resin, plastic, and ceramic containers of the same size, all filled with 400~mL sucrose solution at concentrations from 10\% to 35\%. We also tested resin containers with wall thicknesses of 1, 2, and 3~mm. Figure~\ref{fig:material-thickness} shows that both material and thickness shift the amplitude--viscosity curve.

This result does not contradict the model, and it clarifies the role of calibration. Different containers change the effective stiffness and structural transmission path, much as reinforcement or structural form changes modal behavior in larger mechanical systems \cite{nie2024experimental,huang2021bridge_slab_crack_support_patent,zeng2021anti_overturning_bridge_pier_reinforcement_patent}. The slope remains monotonic, which means the interference can be absorbed into calibrated parameters rather than invalidating the vibration--viscosity relation itself. Therefore, \system treats $k$ as a per-phone, per-container parameter rather than as a universal constant.

\section{System Overview}

\system consists of five modules, as shown in Figure~\ref{fig:system-overview}.

\paragraph{Parameter Calibration Module.}
Different phones and containers exhibit slightly different actuation strengths and structural responses. Before deployment, \system uses several known liquids to calibrate the model parameters. This is calibration of the sensing stack, not liquid-specific classifier training.

\paragraph{Motion Detection Module.}
Device motion corrupts the vibration response and can be easily mistaken for liquid-induced variation. We therefore use Apple's \texttt{CMMotionActivityManager} API \cite{motionAPI} to reject unstable operating conditions. The measurement proceeds only when the phone is stationary enough for the reduced-order model to hold.

\paragraph{Signal Reconstruction Module.}
Because the API sampling rate is lower than the motor frequency, \system first applies SRR to exploit phase diversity across repeated bursts. The resulting waveform remains coarse, so we refine it using OMP-based sparse reconstruction.

\paragraph{Interference Cancellation Module.}
The reconstructed spectrum is used to estimate liquid volume and apply a corresponding volume-compensation vector. We then subtract a calibrated standard straight-path interference template to suppress the direct motor-to-accelerometer leakage.

\paragraph{Viscosity Measurement Module.}
Finally, \system converts the compensated waveform back to the time domain, estimates the steady-state amplitude and decay factor, and calculates viscosity through the model in Section~2.

\section{System Design of \system}

This section explains how we realize the measurement model on a smartphone despite constraints on actuation, sensing, and interference.

\subsection{Vibro-motor Selection}

Not all mobile-phone vibration motors are suitable for viscosity measurement. Our model requires a stable, repeatable excitation with a well-defined dominant frequency. Linear resonant actuators (LRAs) meet this requirement much better than broadband, less-reliable eccentric rotating mass motors. Compact electromagnetic actuation designs also emphasize the value of stable resonant excitation in small sealed devices \cite{li2018electromagnetic_device_patent}. We therefore implement \system on phones that use the iPhone-style LRA, including iPhone~7, 7~Plus, 8, 8~Plus, X, and XR.

A second practical constraint comes from the mobile operating system. In iOS, the motor can vibrate continuously for about 0.5~s and is then forcibly paused for at least 0.25~s. Rather than fighting this behavior, \system embraces burst excitation: steady-state estimation is performed within the 0.5~s active window, and damping is measured during the pause.

\begin{figure}[tbp]
\centering
\includegraphics[width=0.55\linewidth]{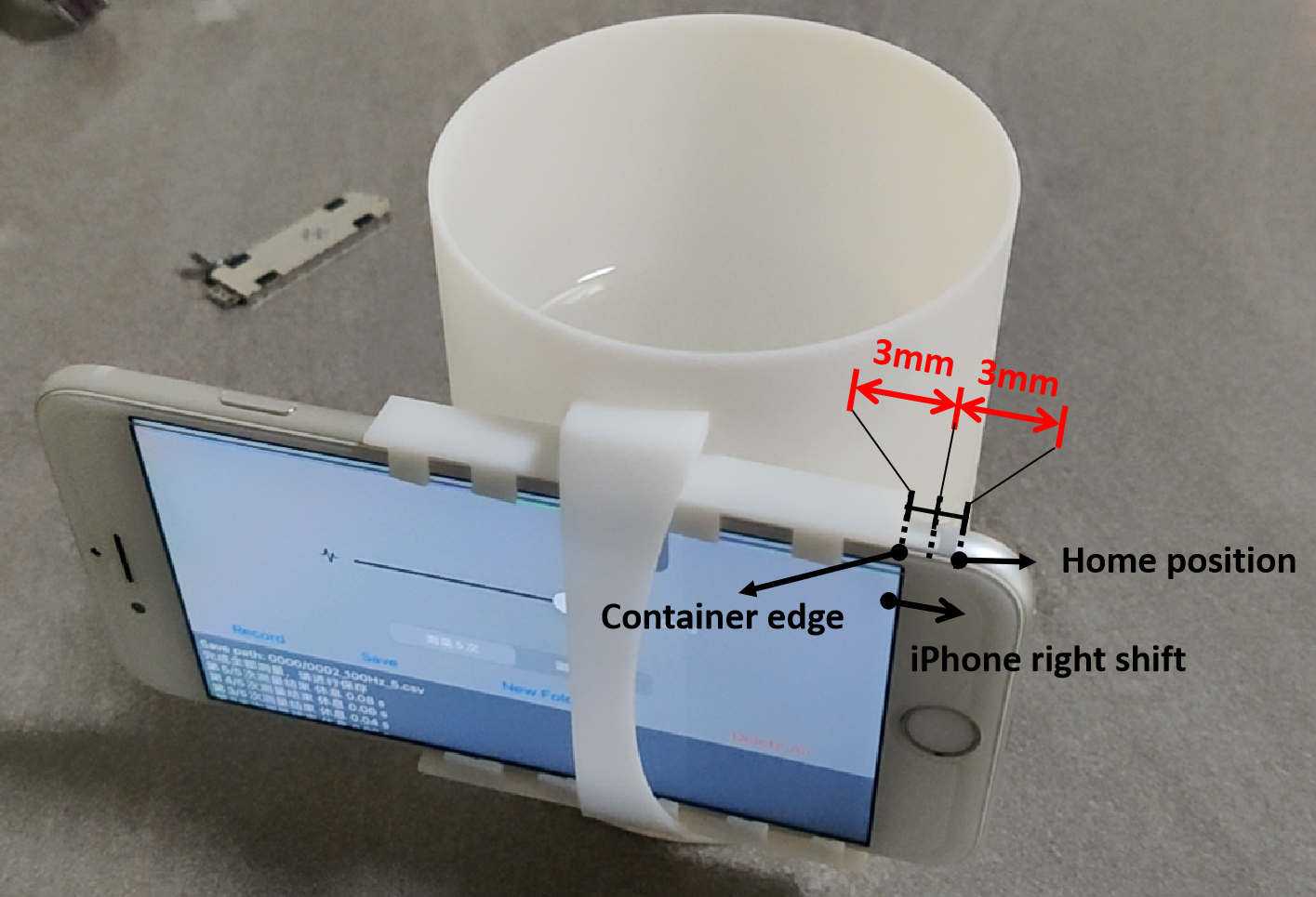}
\caption{Initial phone placement and direction.}
\label{fig:phone-position}
\end{figure}

\begin{figure}[tbp]
\centering
\includegraphics[width=0.65\linewidth]{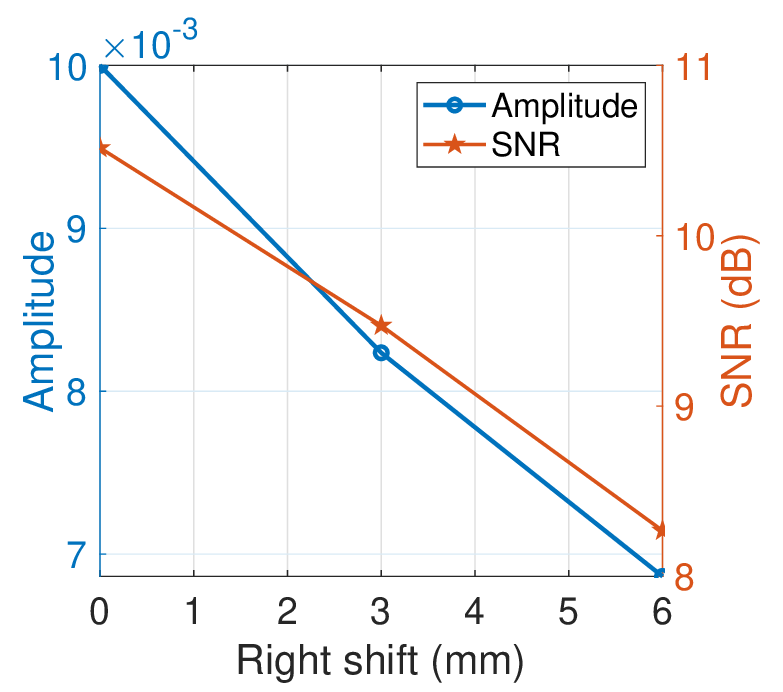}
\caption{Amplitude and SNR at different phone placements.}
\label{fig:amplitude-SNR}
\end{figure}

\begin{figure}[tbp]
\centering
\includegraphics[width=0.65\linewidth]{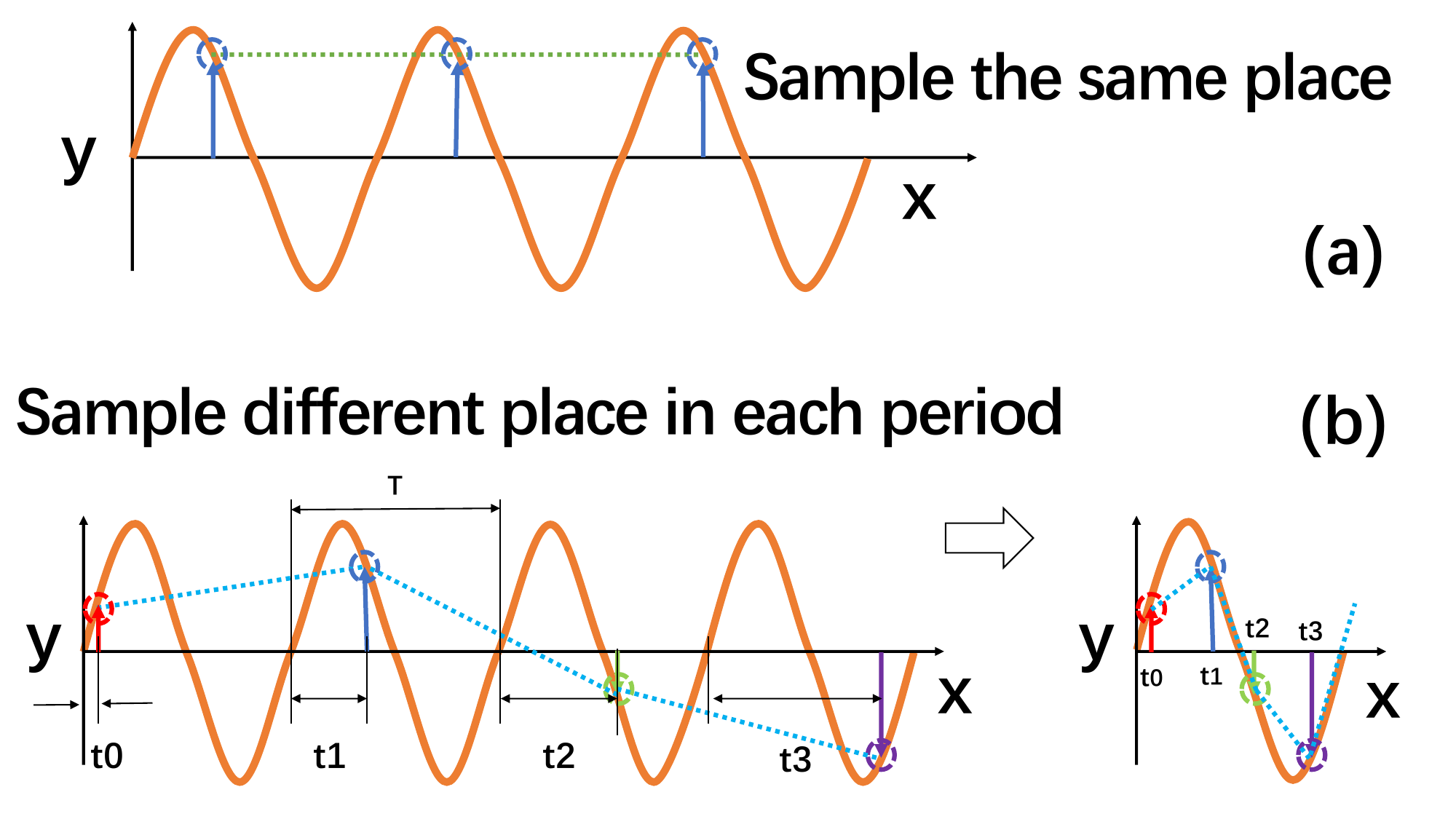}
\caption{Undersampling problem and its solution. (a) Sampling the same phase repeatedly. (b) Phase-diverse sampling across periods.}
\label{fig:undersample-solution}
\end{figure}

\subsection{Mobile Phone Placement Selection}

The phone must be placed so that the motor efficiently drives the container while the container remains mechanically stable. We define the initial placement as the configuration in which the midpoint of the phone's Home button aligns with the edge of the container slot, as shown in Figure~\ref{fig:phone-position}. This placement keeps the motor close to the container while avoiding tilt caused by moving the phone too far beyond the cup's center of support.

We then shifted the phone to the right by controlled offsets and measured the signal-to-noise ratio (SNR) and response amplitude. Figure~\ref{fig:amplitude-SNR} shows that both metrics degrade as the phone moves away from the initial position. This is physically sensible: changing the phone position changes both the torque applied by the motor and the coupling path into the container. We therefore fix the phone placement throughout all experiments so that the motor center remains aligned with the slot edge.

\subsection{Combating Straight Path Interference}

The feasibility study in Section~3 uses separate hardware and therefore does not suffer from the strongest smartphone-specific confounder: the motor can couple directly to the built-in accelerometer through the phone chassis, bypassing the liquid altogether. This straight-path interference dominates the weak liquid-reflected component. The challenge resembles other leakage-channel systems in which a strong direct path masks a weaker informative path \cite{qianru2022magear,liao2024eavesdropping}, and it also resembles overlapping-signal separation problems \cite{jiayi2024chameleon,wang2023multi_keyboard_keystroke_audio_patent}.

To study the direct path at high resolution, we temporarily replaced the phone's built-in accelerometer with a standalone BMI160 and sampled at 1600~Hz. Figure~\ref{fig:transient-signal} shows a clear startup transient during roughly the first 200~ms, so \system discards that interval in every burst.

We then suspended the phone in the air with a soft butterfly clamp so that reflections were strongly attenuated and the recorded signal mainly represented the direct path. Thirty such recordings were segmented into six 50~ms windows, transformed to the frequency domain, and aggregated into a \emph{Standard Straight Path Interference} (SSPI) template. During normal sensing, we subtract a scaled version of this template:
\[
\hat{X}_{clean}(f) = \hat{X}_{meas}(f) - \alpha \hat{X}_{SSPI}(f),
\]
where $\alpha$ is estimated from the dominant motor-frequency bins. After inverse transformation, the remaining waveform is much more sensitive to the liquid-reflected component.

\subsection{\emph{Supersampling Rate Reconstruction} (SRR)}

The iPhone~7 accelerometer hardware can sample up to 1600~Hz, but the public API exposes only 100~Hz, far below the 167~Hz motor frequency. Direct sampling therefore violates the Nyquist criterion and distorts the waveform, especially its peaks. Since our estimator depends on the steady-state amplitude, the distortion must be corrected.

Before applying SRR, we first check that the signal is stable enough across repeated bursts. We align 1600~Hz and 100~Hz recordings on the same time axis and compute the variance after optimal temporal shifting. The difference is only 1.42\%, which indicates that burst-to-burst phase-consistent reconstruction is feasible.

\subsubsection*{Basic Idea of SRR}

Figure~\ref{fig:undersample-solution}(a) illustrates the problem: if a stable periodic signal is sampled too slowly and always with nearly the same phase, the samples fail to trace the true waveform. SRR solves this by introducing a small wait-time offset between repeated bursts. Each burst is therefore sampled at a different phase, as shown in Figure~\ref{fig:undersample-solution}(b). After several bursts, the samples are folded into a single motor period according to their timestamps and sorted by phase. The resulting composite waveform is equivalent to a much higher sampling rate.

\subsubsection*{Why the Sampling Interval Matters}

\begin{figure}[tbp]
\centering
\includegraphics[width=\linewidth]{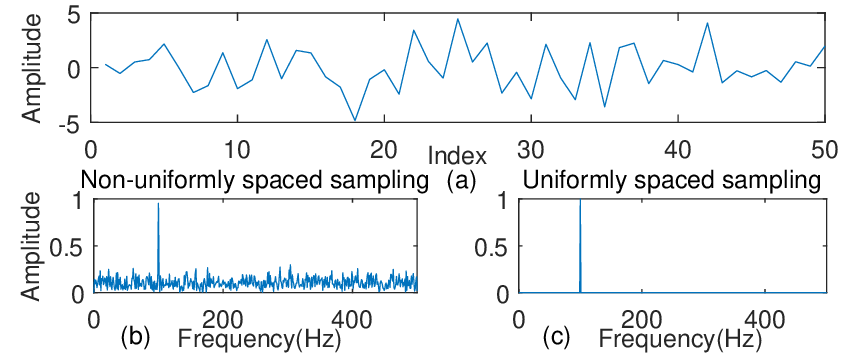}
\caption{The original signal and the magnitude spectrum obtained by sampling it at different time intervals.}
\label{fig:different-time-intervals}
\end{figure}

For a signal sampled at non-uniform times $t_n$, the discrete-time Fourier transform is
\begin{equation}
X\!\left(e^{j\omega}\right)=\sum_{n=1}^{N} x[t_n]e^{-j\omega t_n}.
\label{fourier}
\end{equation}
If the folded phases are poorly distributed, Equation~\ref{fourier} produces strong spectral artifacts. If the timestamps are approximately uniformly distributed over the period, however, the expected spectrum approaches the true continuous-time spectrum:
\[
\mathbb{E}\!\left[X\!\left(e^{j\omega}\right)\right]
= \frac{N}{T_{\max}}\int_0^{T_{\max}} x(t)e^{-j\omega t}\,\mathrm{d}t.
\]
Figure~\ref{fig:different-time-intervals} visualizes this principle: non-uniform phase coverage produces a noisy spectrum, while near-uniform coverage recovers the dominant spectral peak.

In practice, each acquisition cycle contains 0.5~s of active vibration followed by a pause of $0.5$~s $+ n_{th} t_{wait}$, where $n_{th}$ is the burst index and the extra wait time shifts the phase of the next burst. We set $t_{wait}=2.5$~ms and use four bursts, which provides enough phase diversity for stable reconstruction.

\begin{figure}[tbp]
\centering
\includegraphics[width=\linewidth]{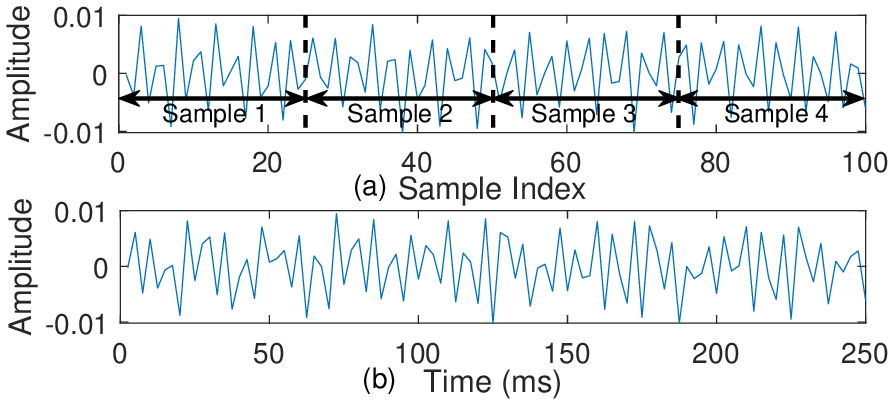}
\caption{Reconstructing the vibration signal with \emph{Supersampling Rate Reconstruction}.}
\label{fig:supersample}
\end{figure}

Figure~\ref{fig:supersample}(a) shows that the raw 100~Hz signal is severely distorted. After SRR across four bursts, Figure~\ref{fig:supersample}(b) recovers a much cleaner waveform whose phase progression resembles the true vibration.

\begin{figure}[tbp]
\centering
\includegraphics[width=\linewidth]{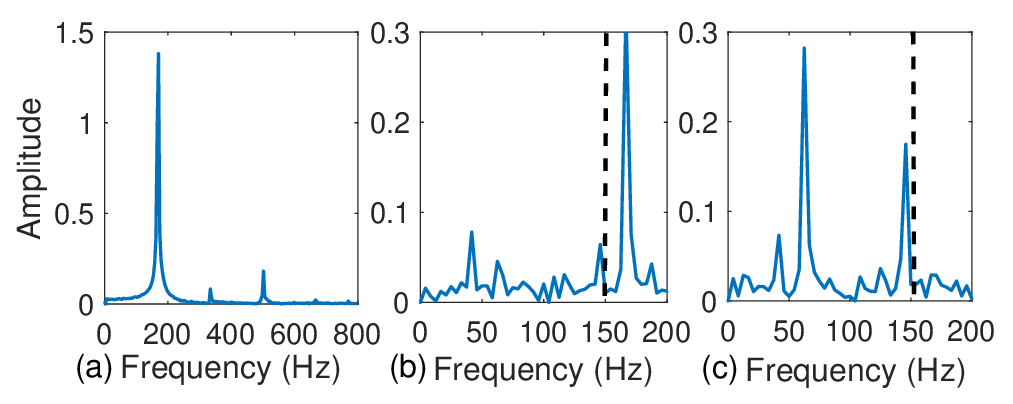}
\caption{Amplitude--frequency curves of the vibration signal. (a) Original high-rate signal from BMI160. (b) Correctly re-ordered signal. (c) Incorrectly re-ordered signal.}
\label{fig:amplitude-frequency-curve}
\end{figure}

\begin{figure}[tbp]
\centering
\includegraphics[width=0.8\linewidth]{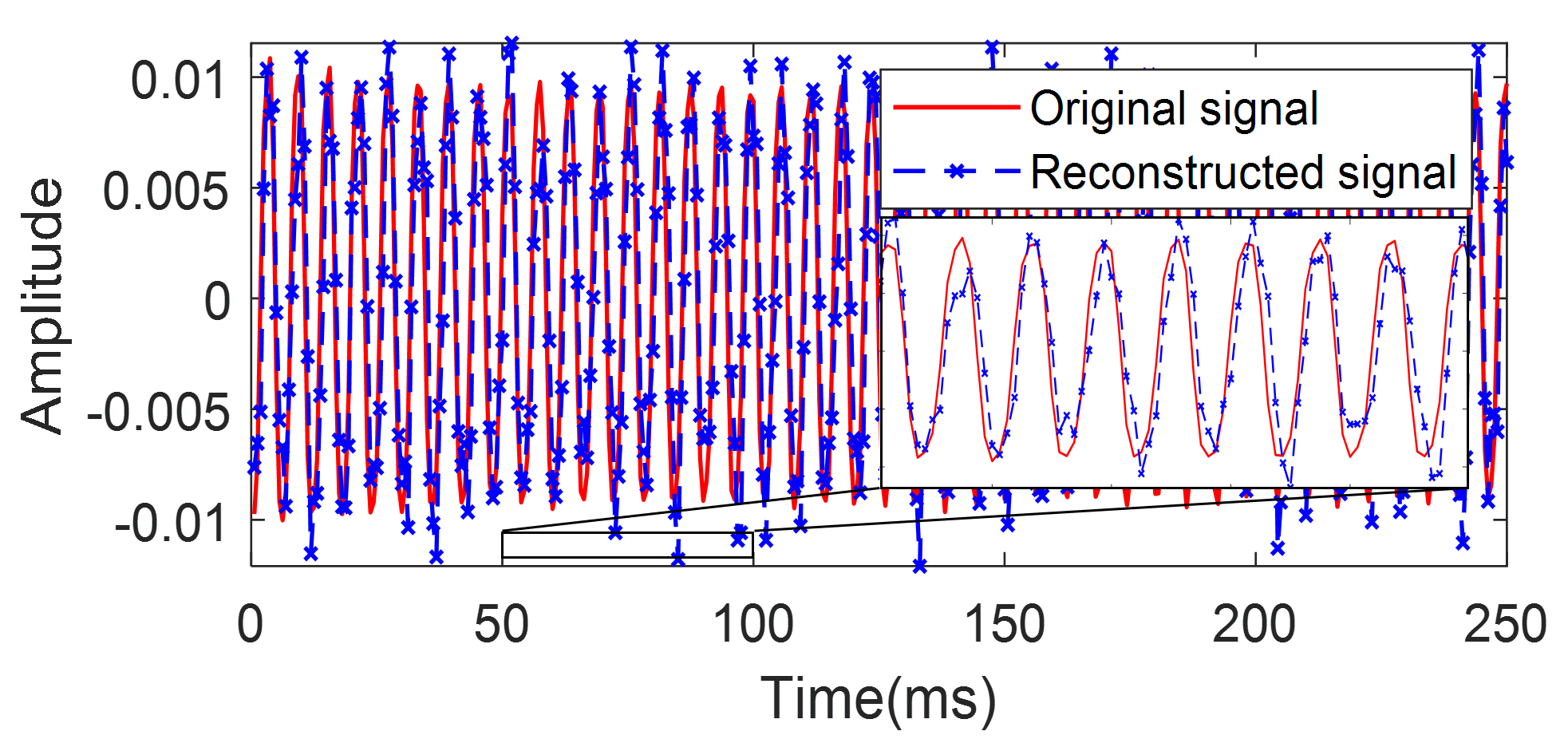}
\caption{Comparison between the reconstructed signal and the original high-rate signal.}
\label{fig:new-supersample}
\end{figure}

\subsubsection*{Choosing the Correct Re-ordering}

One subtle issue remains. When the phone starts the motor, the electrical phase can flip, so the same burst may appear with the opposite sign. Re-ordering $n$ bursts therefore yields up to $2^n$ sign combinations. Figure~\ref{fig:amplitude-frequency-curve} shows how to select the correct one: the right re-ordering concentrates most energy near the motor frequency, producing a dominant peak above 150~Hz that is more than twice the next largest peak. Wrong re-orderings spread energy across low-frequency artifacts because inconsistent sign assignment slows the apparent waveform. \system uses this criterion to automatically choose a valid re-ordering.

\subsection{Orthogonal Matching Pursuit Based Reconstruction (OMPR)}

Even after SRR, the recovered waveform is still relatively coarse. Using its sampled peaks directly would require more bursts and longer latency. We therefore refine SRR with an OMP-based sparse reconstruction stage.

The key observation is that the reconstructed vibration signal is sparse in the frequency domain, as seen in Figure~\ref{fig:amplitude-frequency-curve}. After thresholding small components, only a few dominant frequencies remain. This justifies a compressive-sensing formulation \cite{Candes2008An,tropp2007signal}. Let $\mathbf{x}\in\mathbb{R}^n$ denote the high-rate signal and $\mathbf{y}\in\mathbb{R}^m$ the low-rate observation ($m\ll n$). Then
\[
\mathbf{y} = \bm{\varphi}\mathbf{x}.
\]
If $\mathbf{x}=\bm{\phi}\mathbf{f}$, where $\mathbf{f}$ is sparse in the inverse discrete Fourier basis $\bm{\phi}$, then
\begin{equation}
\mathbf{y} = \bm{\varphi}\bm{\phi}\mathbf{f},
\label{yf}
\end{equation}
where $\bm{\varphi}\bm{\phi}$ is the observation matrix.

In practice, ideal timing matrices are not directly available because the phone adds hardware and API jitter. We therefore estimate an empirical observation matrix using 500 paired recordings: for each sucrose sample, we record the 1600~Hz signal with the standalone accelerometer and the 100~Hz signal with the smartphone API. The latter is first lifted to 167~Hz by SRR, and the paired signals are then used to estimate the observation matrix in Equation~\ref{yf}. Given the learned matrix, OMP iteratively selects the atoms that best explain $\mathbf{y}$ and recovers a high-rate sparse spectrum. Figure~\ref{fig:new-supersample} shows that applying OMP after SRR reconstructs the waveform much more faithfully than applying OMP directly to the 100~Hz samples. The order matters: SRR resolves phase coverage, and OMP then sharpens the sparse spectral structure.

\begin{figure}[tbp]
\centering
\includegraphics[width=\linewidth]{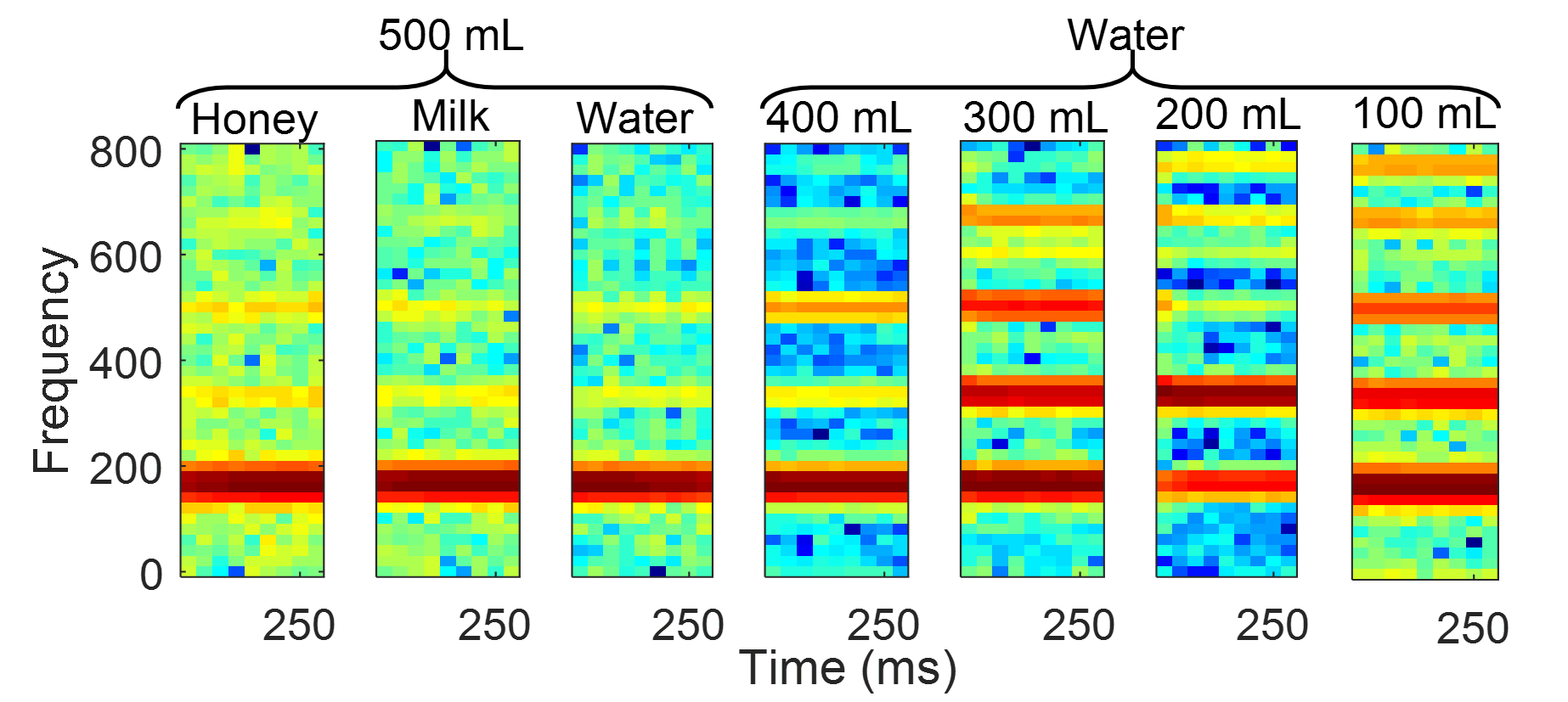}
\caption{Magnitude spectra of liquids with different volumes and different viscosities.}
\label{fig:volume-amplitude-frequency diagram}
\end{figure}

\subsection{Combating Volume Change Impact}

Section~3.4 showed that the same liquid exhibits different vibration signatures at different fill levels. Without compensation, the same liquid would appear to have different viscosities simply because the user poured a different volume.

\subsubsection*{Volume Weight Vector}

For a fixed container, increasing volume adds hydrodynamic loading to the coupled wall--liquid system. A useful reduced-order model is
\[
f_r(V_l) \approx \frac{1}{2\pi}\sqrt{\frac{k_c}{m_c+\alpha V_l}},
\]
where $k_c$ is the effective container stiffness, $m_c$ is the base structural mass, and $\alpha V_l$ models the volume-dependent added mass. As $V_l$ increases, the resonance shifts downward and the frequency-domain envelope changes. This agrees with Figure~\ref{fig:volume-amplitude-frequency diagram}: liquids with different viscosities but the \emph{same} volume have broadly similar left-half spectra, while liquids with the \emph{same} viscosity but different volumes exhibit substantially different spectral envelopes.

We use 500~mL as the reference volume. Let ${a^{ref}}_i$ denote the amplitude of the reference spectrum at frequency bin $i$, and let ${a^{vol}}_i$ denote the amplitude at the same bin for a different volume. The volume weight vector is defined as
\[
W_{volume} =
\left[
\frac{{a^{vol}}_1}{{a^{ref}}_1},
\dots,
\frac{{a^{vol}}_i}{{a^{ref}}_i},
\dots,
\frac{{a^{vol}}_{800\text{Hz}}}{{a^{ref}}_{800\text{Hz}}}
\right].
\]
To normalize a measured spectrum, \system divides each bin by the corresponding element of $W_{volume}$ so that the compensated spectrum matches the 500~mL reference envelope.

\subsubsection*{Liquid Volume Estimation}

Volume compensation requires knowing the volume first. Rather than relying on a closed-form cavity model, \system uses the monotonic frequency shift of the coupled response. It collects 4~s of data, reconstructs the waveform through SRR and OMP, computes the magnitude spectrum, and matches the dominant resonance pattern against a database collected during calibration. Because the relation between volume and resonance is monotonic for a fixed container, this lookup is robust and inexpensive.

\begin{figure}[tbp]
\centering
\includegraphics[width=\linewidth]{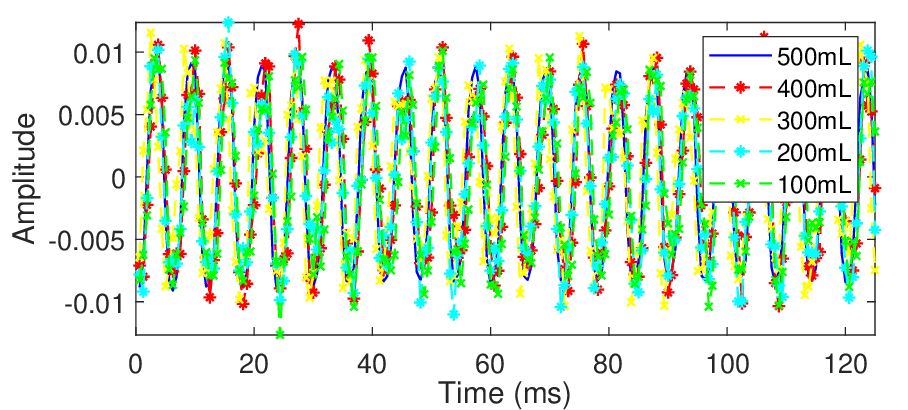}
\caption{Using super-resolution reconstruction to align signals from different volumes.}
\label{fig:volume-super-resolution}
\end{figure}

After applying the estimated volume weights, the reconstructed signals from different volumes become very similar, as shown in Figure~\ref{fig:volume-super-resolution}. The least-squares residual between the aligned waveforms is only 0.023, which indicates that most volume-induced distortion has been removed before viscosity estimation.

\subsection{System Calibration}

System parameters vary across phones and containers. We therefore calibrate four quantities: the effective stiffness $k$, effective mass $m$, motor-force amplitude $f_0$, and geometric constant $\gamma=d/(S_0v)$. We use four liquids with known viscosity---distilled water and 5\%, 10\%, and 20\% sucrose solution---and measure each at four volumes from 100~mL to 500~mL. This gives 16 calibration points.

Let $E(A_i,\Lambda_i;k,m,f_0,\gamma)$ denote the closed-form viscosity estimator from Section~2 applied to calibration sample $i$. Instead of solving only a minimally determined set of simultaneous equations, we fit the parameters by least squares:
\[
(\hat{k},\hat{m},\hat{f}_0,\hat{\gamma}) =
\arg\min_{k,m,f_0,\gamma}
\sum_{i=1}^{16}
\left(
\eta_i - E(A_i,\Lambda_i;k,m,f_0,\gamma)
\right)^2.
\]
This overdetermined calibration is more robust to sensor noise and container-specific transfer differences. Importantly, it remains lightweight: it calibrates the sensing pipeline, not a classifier for liquid identities.

\begin{table}[tbp]
\small
\centering
\caption{Measuring viscosity for various liquids with \system and a viscometer ($\pm$ indicates standard deviation, and GT denotes ground truth).}
\label{tab:liquid_result}
\renewcommand{\arraystretch}{1.05}
\begin{tabular}{cccccc}
\toprule
Liquid & $\beta$ & $f_{\tau}~(\mathrm{N})$ & \system (cP) & GT (cP) & Error (\%) \\
\midrule
5\% salt solution & 0.8775 & 0.0134 & 1.03$\pm$0.019 & 1.00 & 3.00 \\
10\% salt solution & 0.8757 & 0.0142 & 1.09$\pm$0.019 & 1.07 & 1.87 \\
5\% sucrose solution & 0.8760 & 0.0140 & 1.08$\pm$0.025 & 1.06 & 1.89 \\
10\% sucrose solution & 0.8745 & 0.0155 & 1.19$\pm$0.031 & 1.16 & 2.59 \\
Coffee (10g sugar) & 1.3770 & 0.0498 & 3.83$\pm$0.194 & 3.78 & 1.32 \\
Coffee (15g sugar) & 1.4392 & 0.0525 & 4.04$\pm$0.094 & 3.94 & 2.53 \\
Coffee (sugar-free) & 1.3566 & 0.0489 & 3.76$\pm$0.169 & 3.65 & 3.01 \\
Whole milk & 0.8980 & 0.0218 & 1.68$\pm$0.119 & 1.75 & 4.00 \\
Skim milk & 0.8745 & 0.0156 & 1.20$\pm$0.075 & 1.26 & 4.76 \\
Yogurt & 55.545 & 2.0887 & 159.14$\pm$4.800 & 152.45 & 4.39 \\
Beer & 0.8748 & 0.0148 & 1.14$\pm$0.013 & 1.11 & 2.70 \\
Chinese liquor & 0.8745 & 0.0155 & 1.19$\pm$0.031 & 1.23 & 3.25 \\
Coca-Cola & 0.8747 & 0.0149 & 1.15$\pm$0.019 & 1.13 & 1.77 \\
Pepsi-Cola & 0.8745 & 0.0159 & 1.22$\pm$0.013 & 1.24 & 1.61 \\
Chocolate liquor & 14.888 & 0.5625 & 43.28$\pm$0.756 & 40.06 & 3.22 \\
Honey & 1009.8 & 37.899 & 2815.28$\pm$58.88 & 3000.12 & 6.16 \\
Sweet tea & 0.8751 & 0.0166 & 1.28$\pm$0.025 & 1.32 & 3.03 \\
Green tea & 0.8782 & 0.0131 & 1.01$\pm$0.019 & 1.03 & 1.94 \\
Pineapple juice & 33.245 & 1.2518 & 96.29$\pm$3.78 & 100.02 & 3.72 \\
Tomato juice & 28.015 & 1.0555 & 81.19$\pm$1.45 & 79.03 & 2.73 \\
Vinegar & 0.9393 & 0.0261 & 2.01$\pm$0.056 & 2.05 & 1.95 \\
Soy sauce & 0.9799 & 0.0292 & 2.25$\pm$0.113 & 2.18 & 3.37 \\
Soya bean oil & 21.145 & 0.7976 & 61.35$\pm$0.756 & 59.29 & 3.47 \\
Vegetable oil & 10.214 & 0.3870 & 29.77$\pm$0.644 & 30.94 & 3.78 \\
Lard oil & 19.233 & 0.7258 & 55.83$\pm$0.756 & 53.18 & 4.98 \\
Oil (light) & 39.047 & 1.4695 & 113.04$\pm$1.906 & 108.49 & 4.19 \\
Oil (heavy) & 231.24 & 8.6821 & 684.31$\pm$3.325 & 658.12 & 3.98 \\
Disinfected alcohol & 0.8809 & 0.0189 & 1.45$\pm$0.075 & 1.42 & 2.11 \\
Glycerol & 273.09 & 10.253 & 788.67$\pm$4.800 & 800.45 & 1.47 \\
Laundry detergent & 70.319 & 2.6432 & 203.32$\pm$2.331 & 201.05 & 1.13 \\
\bottomrule
\end{tabular}
\end{table}

\section{System Evaluation}

\subsection{Experimental Setup}

In the end-to-end experiments, we attach an iPhone~7 to the side slot of the 3D-printed container, as shown in Figure~\ref{fig:phone-detection}. The vibro-motor frequency is 167~Hz and the built-in accelerometer is sampled at 100~Hz through the iOS API. We use $t_{wait}=2.5$~ms and reconstruct the waveform after four motor restarts. Ground truth still comes from the $ATAGO\text{-}VISCO^{TM}\ 895$ viscometer. Each measurement is repeated 10 times, and we report the mean result unless otherwise stated. All other settings match the feasibility study.

\subsection{Liquid Identification Performance}

We first test 30 liquids at 500~mL and compare the estimated viscosity with ground truth. Table~\ref{tab:liquid_result} spans a wide range, from near-water solutions around 1~cP to honey near 3000~cP. The mean relative error is 2.9\%, and 29 of the 30 liquids remain below 5\% error, and honey is the only outlier and is analyzed further in Section~6.6.

\subsubsection*{Baseline Accuracy}

\begin{figure}[tbp]
\centering
\includegraphics[width=\linewidth]{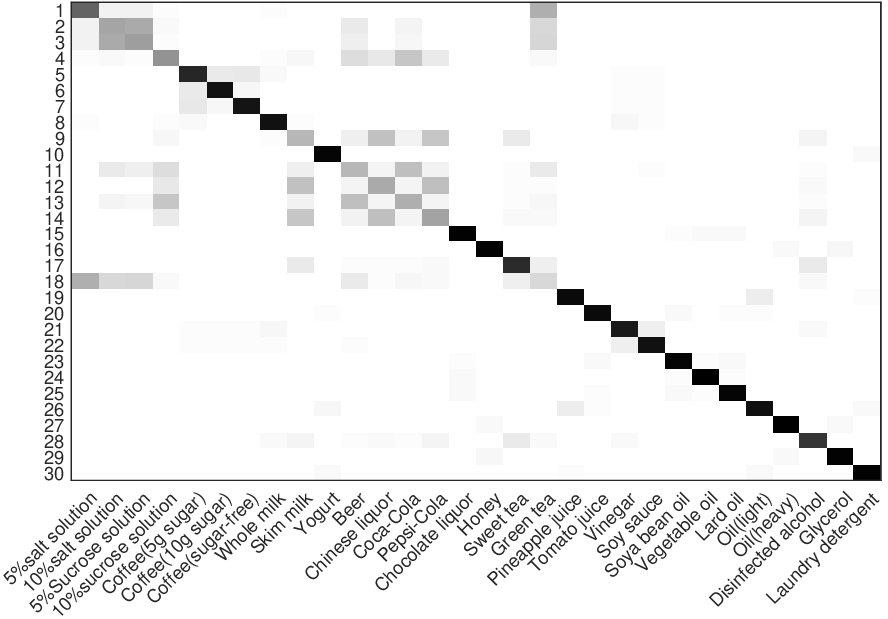}
\caption{Confusion matrix for liquid identification based on the estimated viscosity.}
\label{fig:Confusion-matrix-liquid-identification}
\end{figure}

The most important point in Table~\ref{tab:liquid_result} is that \system performs well even in the hard low-viscosity regime where many liquids are clustered between roughly 1.0 and 1.3~cP. To test how discriminative the estimated viscosity is by itself, we intentionally use a minimal classifier: 1-nearest-neighbor with the scalar estimated viscosity as the only feature. Figure~\ref{fig:Confusion-matrix-liquid-identification} reports an average classification accuracy of 95.47\% over the 30 liquids. This is not a machine-learning contribution, and rather, it shows that the physics-derived viscosity estimate is already highly informative. Most remaining confusion occurs among liquids whose viscosities are extremely close, such as light beverages and dilute solutions.

\subsubsection*{Impact of Straight Path Interference}

\begin{figure}[tbp]
\centering
\includegraphics[width=\linewidth]{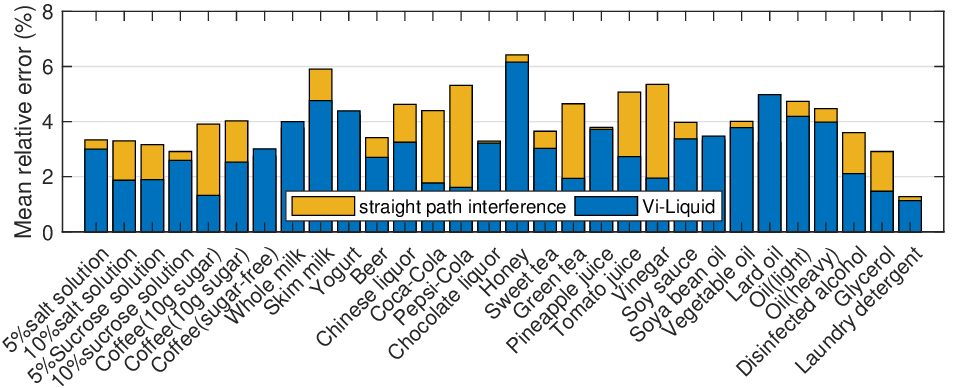}
\caption{Estimated viscosity errors when straight-path interference is not removed.}
\label{fig:straight-path-interference-liquid-identification}
\end{figure}

Figure~\ref{fig:straight-path-interference-liquid-identification} compares the viscosity error before and after straight-path cancellation. Without interference removal, the mean relative error rises to 3.96\%, substantially above the 2.9\% baseline. The comparison confirms that direct motor leakage is not a minor nuisance but a dominant error source that must be modeled explicitly.

\subsubsection*{Impact of Sampling Rate}

\begin{figure}[tbp]
\centering
\includegraphics[width=\linewidth]{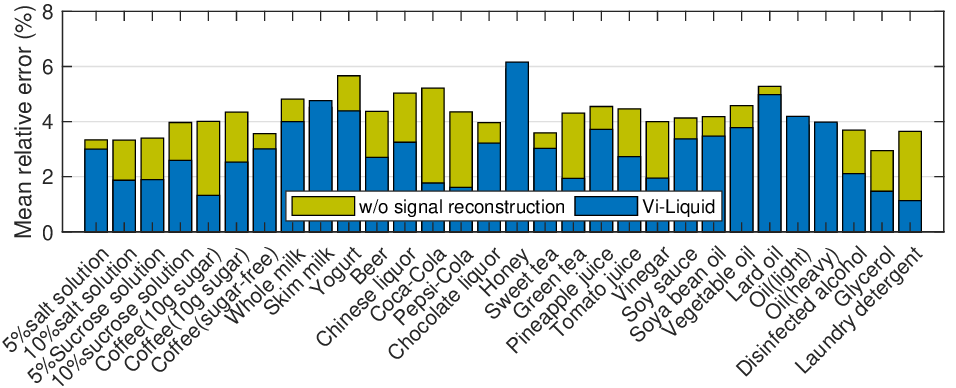}
\caption{Mean errors when the original under-sampled signals are used directly.}
\label{fig:average-errors-original-signals}
\end{figure}

Figure~\ref{fig:average-errors-original-signals} evaluates the effect of signal reconstruction. Using the raw, undersampled signal results in substantial peak errors and increases the mean relative error by 1.41 percentage points. This result matches the analysis in Section~5.4: when the API sampling rate is below the motor frequency, accurate peak amplitude is impossible without phase-aware reconstruction.

\subsubsection*{Impact of Volume}

\begin{figure}[tbp]
\centering
\includegraphics[width=\linewidth]{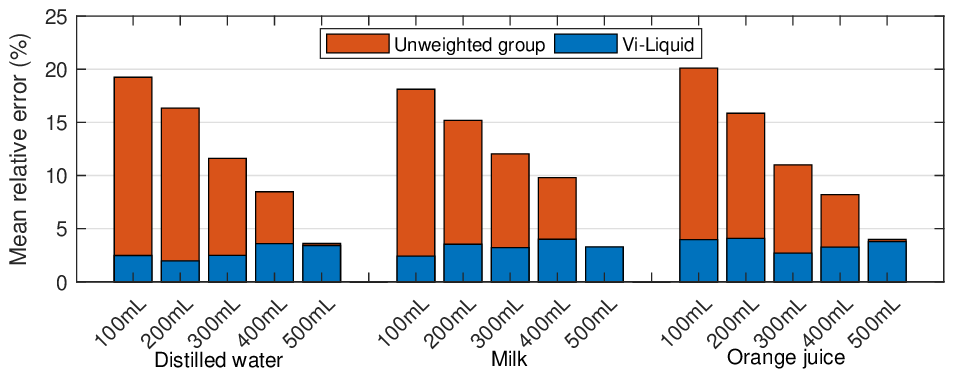}
\caption{Viscosity variation before and after volume compensation.}
\label{fig:before-after-volume-elimination}
\end{figure}

Finally, Figure~\ref{fig:before-after-volume-elimination} tests water, milk, and orange juice at five fill levels from 100~mL to 500~mL. Without compensation, the same liquid yields large and unacceptable estimation variation, and the error can become very large at low fill levels. After frequency-domain volume compensation, the estimated viscosity becomes nearly invariant to volume. This confirms the key lesson from Figure~\ref{fig:volume_attenuation_not_proportional}: volume changes the transfer function, so it must be compensated before viscosity is inferred.

\subsection{Water Contamination Detection}

Non-potable water often differs only subtly from potable water in appearance, especially outside laboratory settings. We therefore test whether \system can discriminate water samples with small viscosity differences: distilled water, tap water, rainwater, puddle water, and water left exposed to air for a long time.

\begin{figure}[tbp]
\centering
\includegraphics[width=0.8\linewidth]{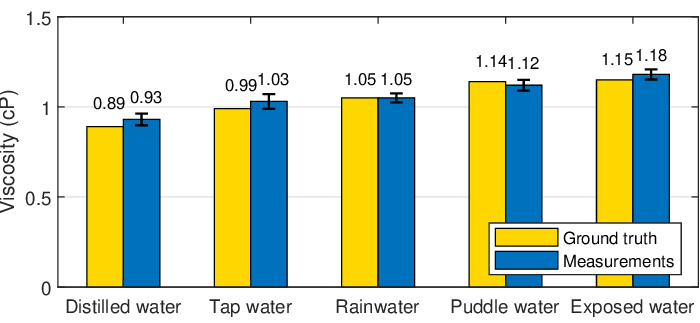}
\caption{Water contamination detection.}
\label{fig:Viscosity-drinkable-non-potable-water}
\end{figure}

Figure~\ref{fig:Viscosity-drinkable-non-potable-water} shows that the measured values preserve the ordering of the five water types, with a mean relative error of 2.56\%. Distilled water and tap water are the closest pair, which makes them the hardest case. Under the controlled temperature and container setting of our experiment, a difference of about 0.1~cP remains measurable. We emphasize that this is a screening-oriented result rather than a regulatory water assay, but it suggests that commodity vibration sensing can reveal small contamination-related changes in fluid properties.

\begin{figure}[tbp]
\centering
\begin{minipage}[t]{\linewidth}
    \centering
    \begin{minipage}[t]{0.32\linewidth}
        \centering
        \includegraphics[width=\linewidth]{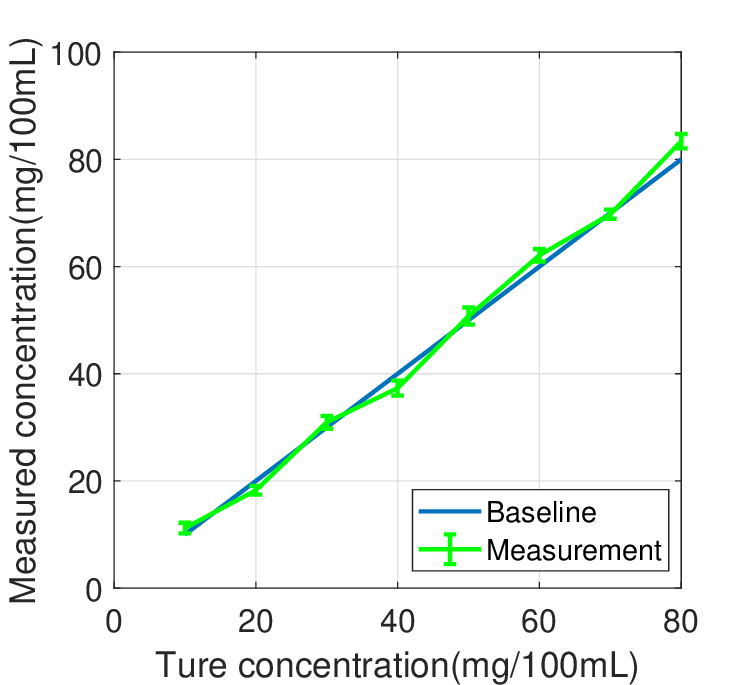}
        \caption*{\footnotesize (a) Sodium urate}
    \end{minipage}
    \hfill
    \begin{minipage}[t]{0.32\linewidth}
        \centering
        \includegraphics[width=\linewidth]{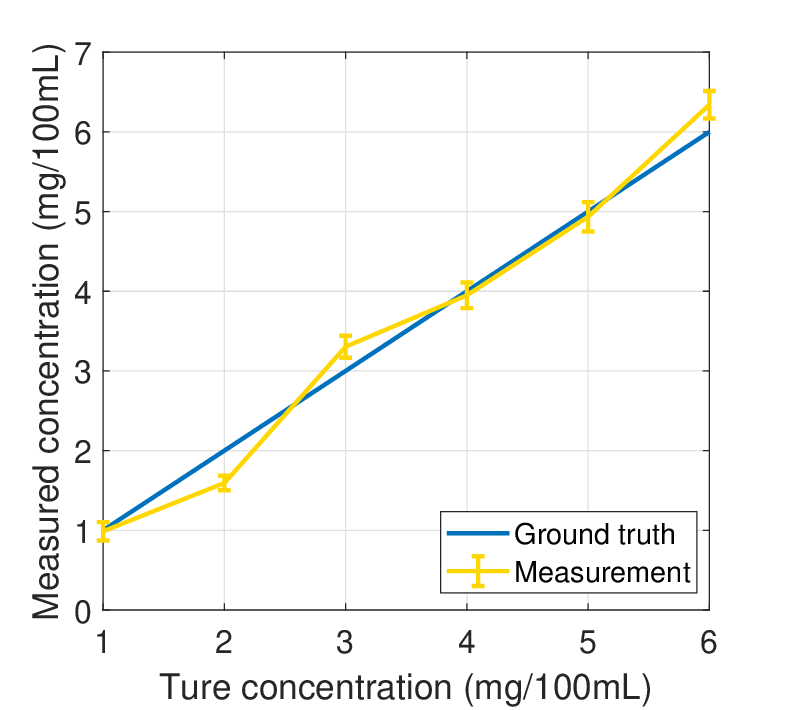}
        \caption*{\footnotesize (b) Ovalbumin}
    \end{minipage}
    \hfill
    \begin{minipage}[t]{0.32\linewidth}
        \centering
        \includegraphics[width=\linewidth]{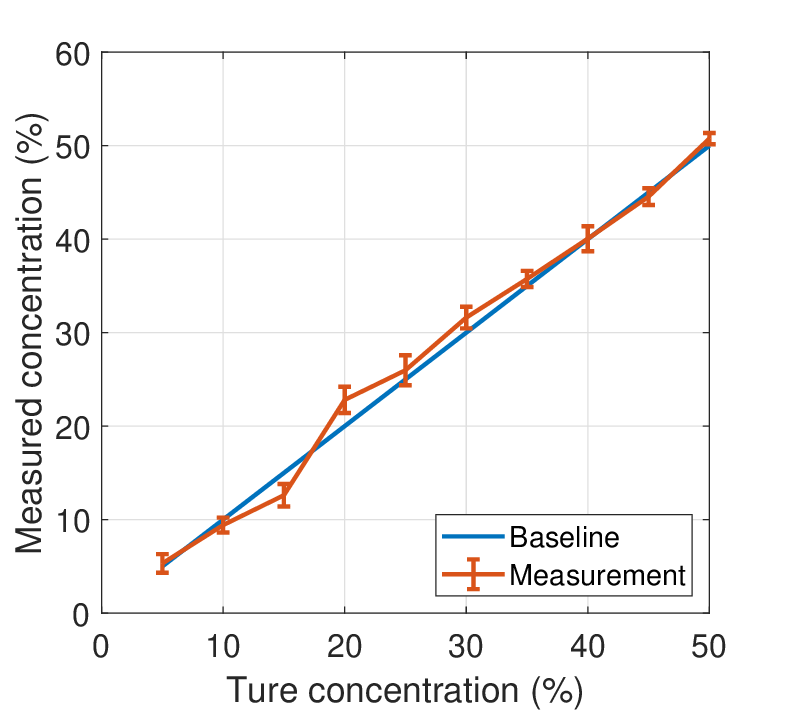}
        \caption*{\footnotesize (c) Alcohol}
    \end{minipage}
\end{minipage}
\caption{Concentration level estimation based on viscosity.}
\label{fig:concentration_level}
\end{figure}

\subsection{Urine Composition Discrimination}

As a proof-of-concept for health-oriented liquid screening, we evaluate whether viscosity can track controlled changes in synthetic urine. The target variables are sodium urate and ovalbumen concentration. These are relevant because abnormal urate and urine protein levels can indicate renal or metabolic problems, and microalbuminuria is a clinically meaningful warning sign \cite{weir2007microalbuminuria}. We stress, however, that this is \emph{not} a clinical diagnostic study: the samples are synthetic, the analytes are controlled, and the system should be understood as a screening-oriented sensor rather than a substitute for laboratory urinalysis.

To generate synthetic urine, we mix distilled water with urea at 38~mg/100~mL, approximating a healthy baseline. We then add controlled amounts of sodium urate and ovalbumen, respectively. Figure~\ref{fig:concentration_level} shows that the estimated concentrations track the ground truth closely. The mean absolute error is 1.15~mg/100~mL for sodium urate and 0.20~mg/100~mL for protein. More broadly, this kind of low-cost longitudinal monitoring aligns with a wider movement toward continuous physiological sensing and closed-loop health technology \cite{huang2016characteristics,ma2014effects,mei2025high}.

\subsection{Alcohol Concentration Measurement}

We also test ethanol concentration estimation. Over the concentration range of our experiments, the effective viscosity of the ethanol--water mixture changes monotonically, allowing a reliable calibration curve. Figure~\ref{fig:concentration_level} shows the relationship between measured viscosity and alcohol concentration, and the mean absolute error is 1.38~mass\%. This result is practical because it suggests that a phone can estimate alcohol content without spectroscopy or additional chemical sensors in the tested range.

\subsection{The Boundary of Measurement}

The honey result in Table~\ref{tab:liquid_result} has the largest error, 6.16\%, which suggests that very high viscosity pushes the system toward its sensitivity limit. As viscosity grows, the response becomes more heavily damped and the incremental change in boundary shear becomes harder to observe with the phone's motor and inertial sensor.

\begin{figure}[tbp]
\centering
\includegraphics[width=\linewidth]{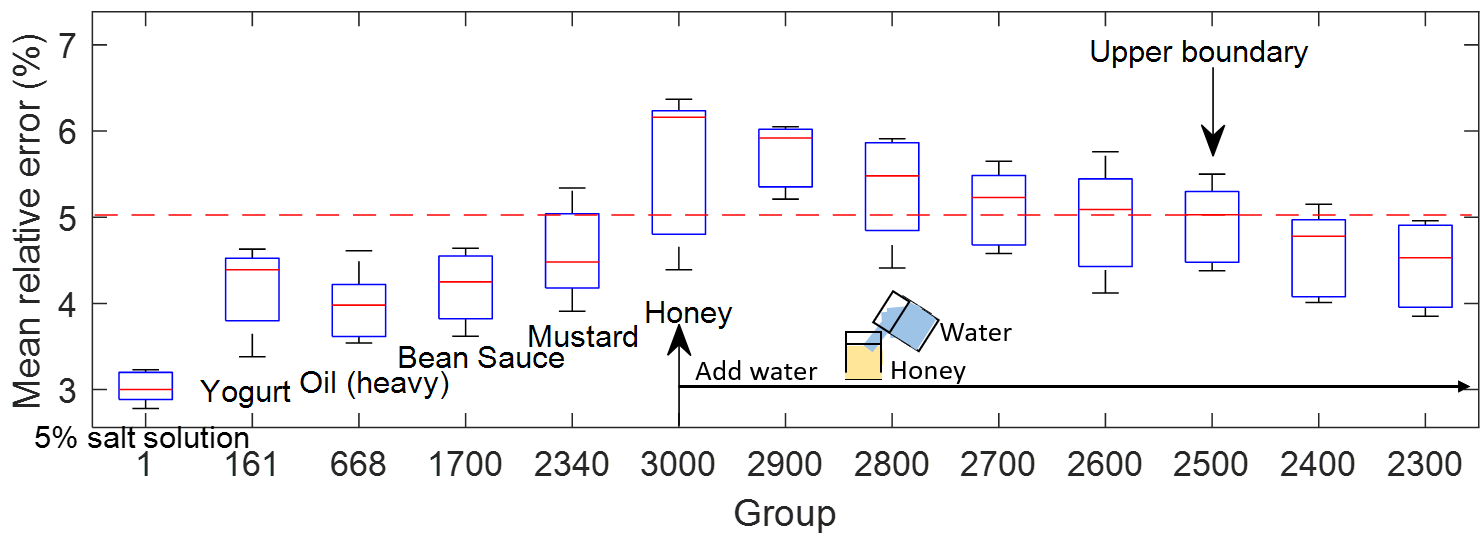}
\caption{Exploring the upper boundary of viscosity measurement.}
\label{fig:Explore-boundary-viscosity}
\end{figure}

We therefore explore the upper operating boundary of \system by progressively diluting honey until the relative error falls back below 5\%. Figure~\ref{fig:Explore-boundary-viscosity} shows that the practical upper limit is around 2500~cP for our hardware and container. This bound is important because it clarifies the regime in which the reduced-order model remains sufficiently observable on a smartphone.

\section{Related Work}

\subsection{Ubiquitous Liquid Testing}

Recent research on ubiquitous liquid testing mainly falls into RF-based and optical/camera-based categories. RF-based systems such as TagScan \cite{wang2017tagscan}, LiquID \cite{dhekne2018liquid}, RFIQ \cite{ha2018learning}, and TagTag \cite{tagtag2019sensys} use RSSI, phase, or permittivity changes to distinguish liquids, but they require specialized radio hardware or instrumented tags. Optical and camera-based systems such as Nutrilyzer \cite{rahman2016nutrilyzer}, Smart-U \cite{huang2018smart}, Al-light \cite{allight2018Ubicomp}, and CapCam \cite{yue2019liquid} exploit absorption, reflection, or surface-wave patterns, but they typically require optical access, transparent media, or dedicated optical components.

Within the broader liquid-sensing line, light-based liquor monitoring \cite{huang2021lili} and beverage deterioration analysis \cite{yongzhi2023beverage} show that commodity devices can sense richer liquid properties when optical channels are available. The preliminary conference paper and later journal version of smartphone vibration-based liquid sensing are reported in \cite{huang2021vi,huang2021portable}. Compared with these systems, \system distinguishes itself by its emphasis on viscosity estimation from inertial sensing alone, without requiring camera access to the liquid or specialized add-on sensors.

\subsection{Measuring Liquid Viscosity}

Viscosity measurement has a long history in fluid mechanics and instrumentation. Classical methods include capillary viscometers \cite{de2004shear}, torsional viscometers \cite{blom1984torsion}, and falling-ball methods \cite{gottlieb1979zero}. Cantilever-based sensors \cite{shih2001simultaneous,riesch2008characterizing,ma2016liquid} improve sensitivity but still require instrumented probes and controlled laboratory setups. Viscosity is also used as an indirect marker in application domains such as seepage analysis \cite{aribudiman2019seepage} and blood rheology \cite{tripolino2017body}.

Non-contact alternatives exist, including ultrasonic surface-wave methods \cite{beyssen2006microfluidic,baird1971noncontact}, photothermal manipulation \cite{motosuke2010noncontact}, and Brillouin-scattering-based optical measurement \cite{xu2003measurement}. These systems can be highly accurate but require hardware far beyond what commodity phones can provide. \system instead accepts a reduced operating range and container-specific calibration in exchange for commodity deployment and zero external sensing hardware.

\subsection{Physics-Driven Mobile Sensing and Interference Handling}

\system also belongs to a larger family of systems that extract fine-grained state from subtle physical couplings. In vibration and acoustic sensing, examples include surface-touch sensing \cite{liu2017vibsense}, bone-conductive text entry \cite{huang2018oinput,yongzhi2018oinput,wu2019bone_conduction_input_patent}, keystroke and tap sensing \cite{huang2018oinput,yongzhi2018oinput,wu2019bone_conduction_input_patent}, bed-based heart and respiration monitoring \cite{monitoringheart2017sensys}, gait authentication \cite{footprintID2017ubicomp}, touch-based authentication \cite{yang2016vibid}, and gesture recognition \cite{viband2016uist,serendipity2016chi}. What these systems share with \system is a reliance on weak but structured mechanical signals rather than explicit chemical measurements.

Recent work has also emphasized interference separation and the exploitation of unconventional channels. Magnetic leakage can reveal audio content \cite{qianru2022magear,liao2024eavesdropping}. Overlapping keystroke signals can be separated adaptively \cite{jiayi2024chameleon,wang2023multi_keyboard_keystroke_audio_patent}, and fine-grained activity can be inferred from lightweight motion signals \cite{wang2021action_recognition_patent,xia2025self}, and and ear-canal or electromagnetic side channels can be reconstructed for subtle physiological or security-related sensing \cite{huang2025reconstructing,liao2025camfirm}. These works reinforce the design philosophy behind \system: when the informative signal is weak, careful physical modeling and signal reconstruction matter more than adding a more complex classifier.

A related direction is the packaging of sensing into practical assistive or consumer form factors. Examples include sensory-substitution methods \cite{kaishun2022auxiliary,wu2020sensory_substitution_assistive_patent}, assistive headphones and glasses \cite{chen2022lisee,huang2023sensory_substitution_blind_assist_glasses_patent,wang5357714collarwad}, and optical consumer products for daily hygiene monitoring \cite{chen2023lit,chen2024optical}. These systems are not liquid sensors, but they demonstrate the same end-to-end challenge faced by \system: turning subtle physical signals into robust, user-facing everyday instrumentation.

\subsection{Deployment, Privacy, and Networked Sensing}

The broader deployment context of \system reaches beyond standalone signal processing. Commodity RF and thermal systems have demonstrated fire detection, humidity sensing, and boiler monitoring \cite{shuxin2017wi,yongzhi2018mm,huang2018mm,wu2018wireless_fire_detection_alarm_patent, wu2018wireless_humidity,wu2018wireless_humidity,duan2018thermal}. More recent work explores integrated sensing and communication over LPWAN and LoRa-style links \cite{wang2025senlora,wu2025low,wu2023cluster_based_lpwan_recovery_patent,wu2023channel_aware_rl_lpwan_isac_patent}. These directions suggest a natural extension of \system toward distributed water or beverage monitoring at building or city scale.

At the same time, long-term deployment raises privacy and security questions. Efficient privacy-preserving federated learning and homomorphic-encryption techniques \cite{xie2024efficiency,qipeng2024litecrypt}, sensor-agnostic telemonitoring \cite{xie2025harmony}, secure protocol design \cite{wang2025attack}, and recent momentum-based federated optimization methods \cite{li2025fedwcm,huang2025fedwmsam,li2025fast} indicate how future versions of \system could update models across many users without centralizing raw sensor traces. Cloud-side visualization and multimodal analysis \cite{liu2023vr_panorama_cloud_transcoding_patent,huang2024label} may further improve remote monitoring and data interpretation. Altogether, these works reflect the broader pervasive-computing agenda represented by recent community efforts \cite{agu2025}: robust physical sensing, practical deployment, and privacy-preserving intelligence should be designed together rather than in isolation.

\section{Conclusion and Future Work}

This paper presents \system, a smartphone-based liquid sensing system that estimates viscosity from active vibration. The central technical idea is simple but powerful: viscosity can be identified only when steady-state amplitude and decay-stage attenuation are jointly modeled. We derive that estimator, validate its assumptions with standalone hardware, and then realize it on a smartphone by solving three practical system problems: severe under-sampling, strong straight-path interference, and volume-dependent resonance shift. Across 30 liquids, the resulting system achieves 2.9\% mean relative viscosity error and 95.47\% candidate-set identification accuracy.

The current system is promising, but several limitations remain.

\paragraph{Container dependence.}
Our experiments use a calibrated container and controlled placement of the phone. Generalizing to arbitrary cups without sidewall attachment will require either better structural self-calibration or alternative coupling strategies.

\paragraph{Temperature sensitivity.}
Viscosity is strongly temperature-dependent, so field deployment will require explicit temperature compensation rather than assuming a stable room environment.

\paragraph{Non-uniqueness of viscosity.}
Viscosity is highly informative but not universally unique. Different liquids can share similar viscosities, especially in complex mixtures. Future systems should fuse additional properties such as density, surface tension, optical absorption, or permittivity. This is a natural direction given recent progress in smartphone optical liquid analysis \cite{huang2021lili,yongzhi2023beverage,chen2023lit,chen2024optical,wu2020structlight} and label-efficient multimodal modeling \cite{huang2024label}.

\paragraph{High-viscosity boundary and health deployment.}
The current hardware saturates around 2500~cP. In addition, our urine experiments are proof-of-concept studies on synthetic samples rather than medical diagnostics. Future health-oriented deployment should therefore combine better hardware sensitivity with privacy-preserving telemonitoring and secure model updates \cite{xie2024efficiency,qipeng2024litecrypt,xie2025harmony,wang2025attack,li2025fedwcm,huang2025fedwmsam,li2025fast}. In parallel, distributed sensing and communication infrastructures \cite{wang2025senlora,wu2025low,wu2023cluster_based_lpwan_recovery_patent,wu2023channel_aware_rl_lpwan_isac_patent} may enable networked water-quality or beverage monitoring.

\section*{Acknowledgments}

This research was supported in part by the China NSFC Grant (U2001207, 61872248), Guangdong NSF 2017A030312008, Shenzhen Science and Technology Foundation (No. ZDSYS20190\\902092853047), and the Project of DEGP (No.~2019KCXTD005). Additional support came from the Guangdong ``Pearl River Talent Recruitment Program'' under Grant 2019ZT08X603, Guangdong Science and Technology Foundation (2019B111103001, 2019B020209001), the China NSFC 61872246, and the Guangdong Special Support Program. Kaishun Wu is the corresponding author.

\bibliographystyle{unsrtnat}
\bibliography{ViLquid1}

\end{document}